\documentclass[aps,prb,twocolumn,superscriptaddress,longbibliography]{revtex4-2}
\usepackage{amsmath,amsthm,verbatim,amssymb,amsfonts,amscd,graphicx}
\usepackage[colorlinks=true,citecolor=blue,linkcolor=magenta,urlcolor=blue]{hyperref}
\usepackage{color}
\usepackage{soul}

\newtheorem{lemma}{Lemma} 

\usepackage{nicematrix}
\makeatletter
\DeclareRobustCommand\vdots{%
  \mathpalette\@vdots{}%
}
\newcommand*{\@vdots}[2]{%
  \sbox0{$#1\cdotp\cdotp\cdotp\m@th$}%
  \sbox2{$#1.\m@th$}%
  \vbox{%
    \dimen@=\wd0 %
    \advance\dimen@ -3\ht2 %
    \kern.5\dimen@
    \dimen@=\wd2 %
    \advance\dimen@ -\ht2 %
    \dimen2=\wd0 %
    \advance\dimen2 -\dimen@
    \vbox to \dimen2{%
      \offinterlineskip
      \copy2 \vfill\copy2 \vfill\copy2 %
    }%
  }%
}
\DeclareRobustCommand\ddots{%
  \mathinner{%
    \mathpalette\@ddots{}%
    \mkern\thinmuskip
  }%
}
\newcommand*{\@ddots}[2]{%
  \sbox0{$#1\cdotp\cdotp\cdotp\m@th$}%
  \sbox2{$#1.\m@th$}%
  \vbox{%
    \dimen@=\wd0 %
    \advance\dimen@ -3\ht2 %
    \kern.5\dimen@
    \dimen@=\wd2 %
    \advance\dimen@ -\ht2 %
    \dimen2=\wd0 %
    \advance\dimen2 -\dimen@
    \vbox to \dimen2{%
      \offinterlineskip
      \hbox{$#1\mathpunct{.}\m@th$}%
      \vfill
      \hbox{$#1\mathpunct{\kern\wd2}\mathpunct{.}\m@th$}%
      \vfill
      \hbox{$#1\mathpunct{\kern\wd2}\mathpunct{\kern\wd2}\mathpunct{.}\m@th$}%
    }%
  }%
}
\makeatother

\newcommand{\ket}[1]{\ensuremath{\left| #1 \right\rangle}}
\newcommand{\bra}[1]{\ensuremath{\left\langle #1 \right|}}
\newcommand{\inp}[2]{\left\langle #1| #2\right\rangle}

\newcommand{\bea}{\begin{equation}\begin{aligned}}
\newcommand{\eea}{\end{aligned}\end{equation}}

\begin{document}
\title{Boundary Perturbation Effects in Quantum Systems with Conserved Energy and Continuous Symmetry}
\author{Qucheng Gao}
\email{\textcolor{magenta}{gaoqc@bc.edu}}
\affiliation{Department of Physics, Boston College, Chestnut Hill, Massachusetts 02467, USA}
\author{Xiao Chen}
\email{\textcolor{magenta}{chenaad@bc.edu}}
\affiliation{Department of Physics, Boston College, Chestnut Hill, Massachusetts 02467, USA}

\begin{abstract}

We investigate one-dimensional systems with both energy conservation and a continuous symmetry, focusing on the impact of a boundary perturbation that breaks the continuous symmetry. Our study reveals two distinct dynamical phases: one in which the corresponding charge exhibits extensive fluctuations, and another where the charge remains conserved. These phases appear in both free and interacting models. We interpret this behavior through a boundary-induced pumping mechanism, which estimates the amplitude connecting two degenerate states from different charge sectors via a local charge–non-conserving operator. In the Floquet setting, we show that the frozen phase can survive at high driving frequencies but vanishes at low frequencies. 
This phenomenon is exact in free-fermion systems in the thermodynamic limit, but in interacting systems it appears only at finite system size.
The emergence of the charge-frozen phase is attributed to effective energy conservation, and we demonstrate that this phase disappears when effective energy conservation is broken or replaced by other symmetries.

\end{abstract}

\maketitle

\tableofcontents

\section{Introduction}
The influence of a single impurity on a many-body system presents a fascinating and fundamental question. One particularly well-known example is the Kondo effect, where a magnetic impurity can dramatically alter the transport properties of a metal at low temperatures~\cite{kondo1964resistance,hewson1997kondo}. This phenomenon has been further interpreted through the lens of boundary conformal field theory~\cite{cardy2004boundary,andrei2020boundary}.

Beyond this, similar setups have been explored in the context of highly non-equilibrium dynamics induced by boundary perturbations. These studies span both classical stochastic processes~\cite{blythe2007nonequilibrium} and quantum systems~\cite{landi2022nonequilibrium}. Recently, such setups have also been investigated from the perspective of quantum information~\cite{gao2024information,gao2024scrambling}. Remarkably, it was found that a single interaction term can induce the scrambling of information across the entire free-fermion random circuit system in low dimensions. However, in three dimensions, tuning the strength of the impurity induced a scrambling phase transition~\cite{gao2024scrambling}.

Building on previous work, this paper investigates a one-dimensional quantum {\it Hamiltonian} system with a global 
$U(1)$ symmetry, which ensures the conservation of the total charge. We add a boundary perturbation to the Hamiltonian that explicitly breaks the 
$U(1)$ symmetry and address the following question: How does the charge in the bulk evolve in the presence of this boundary perturbation? 

At first glance, one might expect the boundary perturbation to allow charge to flow into or out of the system, leading an initial state with a fixed charge to evolve into a superposition of states spanning multiple charge sectors. This expectation holds in systems without energy conservation, such as those governed by driven dynamics or quantum circuit models, where thermalization across the entire Hilbert space can occur. However, in our system, the dynamics are governed by Hamiltonian evolution, which conserves the total energy. As we demonstrate in this paper, the interplay between boundary perturbations and energy conservation results in more intricate behavior. By varying the parameters in the bulk of the system, we uncover two distinct phases: in one phase, charge fluctuations are negligible, and the total charge remains nearly conserved, with only an $O(1)$
change; in the other, significant charge fluctuations arise, allowing the total charge to vary dramatically.

The existence of the charge-frozen phase arises from effective energy conservation, which constrains the free movement of charge into or out of the bulk. This behavior can be simply understood as follows: moving charge into or out of the system requires additional energy, which is forbidden under energy conservation. 
We substantiate these findings with numerical results for free and interacting fermionic systems (Sec.~\ref{sec::2} and Sec.~\ref{sec::3}), and interacting spin systems (Sec.~\ref{sec::4}). To explain the behavior in both the charge-fluctuating and charge-frozen phases, we introduce a pumping mechanism based on the effective Hamiltonian obtained via degenerate perturbation theory (Sec.~\ref{sec::5}). This mechanism is equivalent to evaluating the matrix element of a local charge–non-conserving operator connecting two degenerate states with different charges: it vanishes in the charge-frozen phase and becomes nonzero in the charge-fluctuating phase.
We further study Floquet dynamics and demonstrate that the effective energy conservation is a necessary condition for the existence of the phase transition (Sec.~\ref{sec::6}).
We summarize our results in Sec.~\ref{sec::7}.

\begin{figure*}
    \centering
    \includegraphics[width=0.99\linewidth]{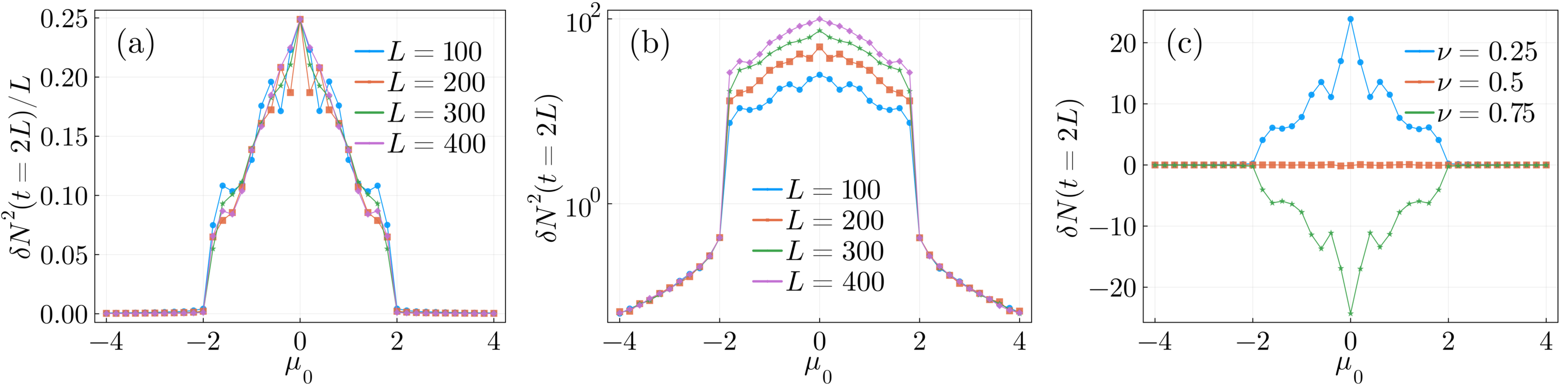}
    \caption{Free-fermion chain. (a) The density of charge variance $\delta N^2/L$ as a function of $\mu_0$ for different system sizes. (b) The charge variance $\delta N^2$ as a function of $\mu_0$ for different system sizes on a log-lin scale. Here, $t_0=\Delta=1$, $\nu=0.5$, and the results are averaged over $1000$ samples. (c) The charge difference $\delta N$ as a function of $\mu_0$ for different filling factors. Note that when the initial state is at half-filling, the charge difference cannot detect the phase transition. Here, $t_0=\Delta=1$, $L=100$, and the results are averaged over $1000$ samples.
 }\label{fig1}
\end{figure*}

\section{Boundary perturbation on a free-fermion chain}\label{sec::2}
We start with a one-dimensional free-fermion chain of finite size $L$ with periodic boundary condition (PBC). The Hamiltonian is
\bea\label{eq::fh0}
\hat{H}_0 &=\sum_{j=1}^L t_0(\hat{c}_j^{\dagger}\hat{c}_{j+1} + \hat{c}_{j+1}^{\dagger}\hat{c}_{j})- \sum_{j=1}^L \mu_0\hat{n}_j\\
&=\sum_k(2t_0\cos k-\mu_0) \hat{c}_k^{\dagger}\hat{c}_{k},
\eea
where $\hat{c}_j$ is the annihilation operator of the fermion on site $j$, $\hat{n}_j\equiv\hat{c}_j^{\dagger}\hat{c}_{j}$, $t_0$ is the hopping parameter, and $\mu_0$ is the chemical potential. The second equality uses the Fourier transform $\hat{c}_j = \frac{1}{\sqrt{L}}\sum_k e^{ikj}\hat{c}_k$. For simplicity, we assume $t_0>0$ throughout. Note that $\hat{H}_0$ commutes with the number operator $\hat{N}=\sum_j \hat{n}_j$ and therefore the total particle number is conserved. We prepare our system in a state $\ket{\Psi(0)}$ with a fixed particle number $N$, i.e., 
\bea
    \hat{N}\ket{\Psi(0)} = N\ket{\Psi(0)}.
\eea
At time $t=0$, we introduce a boundary perturbation $\hat{H}_B$,
\bea\label{eq::fhB}
    \hat{H}_B =& \Delta(\hat{c}_1^{\dagger}\hat{c}_2^{\dagger} + \hat{c}_2\hat{c}_1)\\
    =&\frac{\Delta}{L}\sum_{k_1k_2}\Big(  e^{-i(k_1+2k_2)} \hat{c}_{k_1}^{\dagger}\hat{c}_{k_2}^{\dagger}
    + e^{i(k_1+2k_2)} \hat{c}_{k_2}\hat{c}_{k_1} \Big),
\eea
which does not commute with $\hat{N}$. For times $t>0$, the state of the system evolves as
\bea\label{eq::evolution}
    \ket{\Psi(t)} = e^{-i\hat{H}t}\ket{\Psi(0)},
\eea
where $\hat{H}=\hat{H}_0+\hat{H}_B$. In this setup, the boundary perturbation can disrupt particle number conservation during time evolution. Our goal is to investigate \emph{how the boundary perturbation affects the initially conserved quantity in the bulk}. To quantify the changes in the initially conserved particle number, we analyze the charge variance ~\cite{yu2025symmetry}:
\bea\label{eq:dnt}
    \delta N^2(t) = \bra{\Psi(t)} (\hat{N}-\bra{\Psi(t)} \hat{N} \ket{\Psi(t)})^2 \ket{\Psi(t)}.
\eea

The quadratic Hamiltonian and the Gaussian free-fermionic state allow for efficient simulations on classical computers \cite{bravyi2004lagrangian,ravindranath2025free}.
We prepare a randomly filled product state with a filling factor $\nu$, ensuring that the initial state has $N=L\nu$.
This state is evolved for a long time with 
$t=2L$ (we set $\hbar = 1$ and measure time in units of $1/t_0$), and the steady-state value of $\delta N^2$ is computed as a function of $\mu_0$ (see also Appendix \ref{app::1} for the particle dynamics and energy dynamics). The results for various system sizes $L$ are shown in Fig.~\ref{fig1}(a) and Fig.~\ref{fig1}(b). We observe a phase transition at $|\mu_0|=2t_0$. When $|\mu_0|>2t_0$, the boundary perturbation induces only an $O(1)$ change in the charge in the thermodynamic limit. This is evident in Fig.~\ref{fig1}(b), where $\delta N^2(t=2L)$ remains a small finite value across different system sizes. On the other hand, when $|\mu_0|<2t_0$, there are strong charge fluctuations with $\delta N^2(t=2L)$ proportional to the system size $L$. This behavior indicates that the final state is an extensive superposition of wavefunctions with different charge sectors.
Note that, in the free fermion model, the phase transition point coincides with the gap closing point of the single-particle spectrum. We will give an explanation of this behavior later in Sec.~\ref{sec::5}.
The phase transition can also be observed from the charge difference between the final state and the initial state:
\bea
    \delta N(t) = \bra{\Psi(t)} \hat{N} \ket{\Psi(t)} - \bra{\Psi(0)} \hat{N} \ket{\Psi(0)}.
\eea
As shown in Fig.~\ref{fig1}(c), when $|\mu_0|<2t_0$, the charge undergoes significant changes unless the initial state is exactly at half-filling. In contrast, when $|\mu_0|>2t_0$, the charge changes by at most an 
$O(1)$ constant in the presence of boundary perturbations.  
At half-filling, in the charge-fluctuating phase, although the final state is a superposition of states with different particle numbers, the contributions from states with particle numbers greater than $L/2$ are balanced by those with particle numbers less than $L/2$. As a result, the average particle number remains equal to that of the initial state. Therefore, the charge difference $\delta N$ is not an effective probe at half-filling.

\begin{figure}
    \centering
    \includegraphics[width=0.99\linewidth]{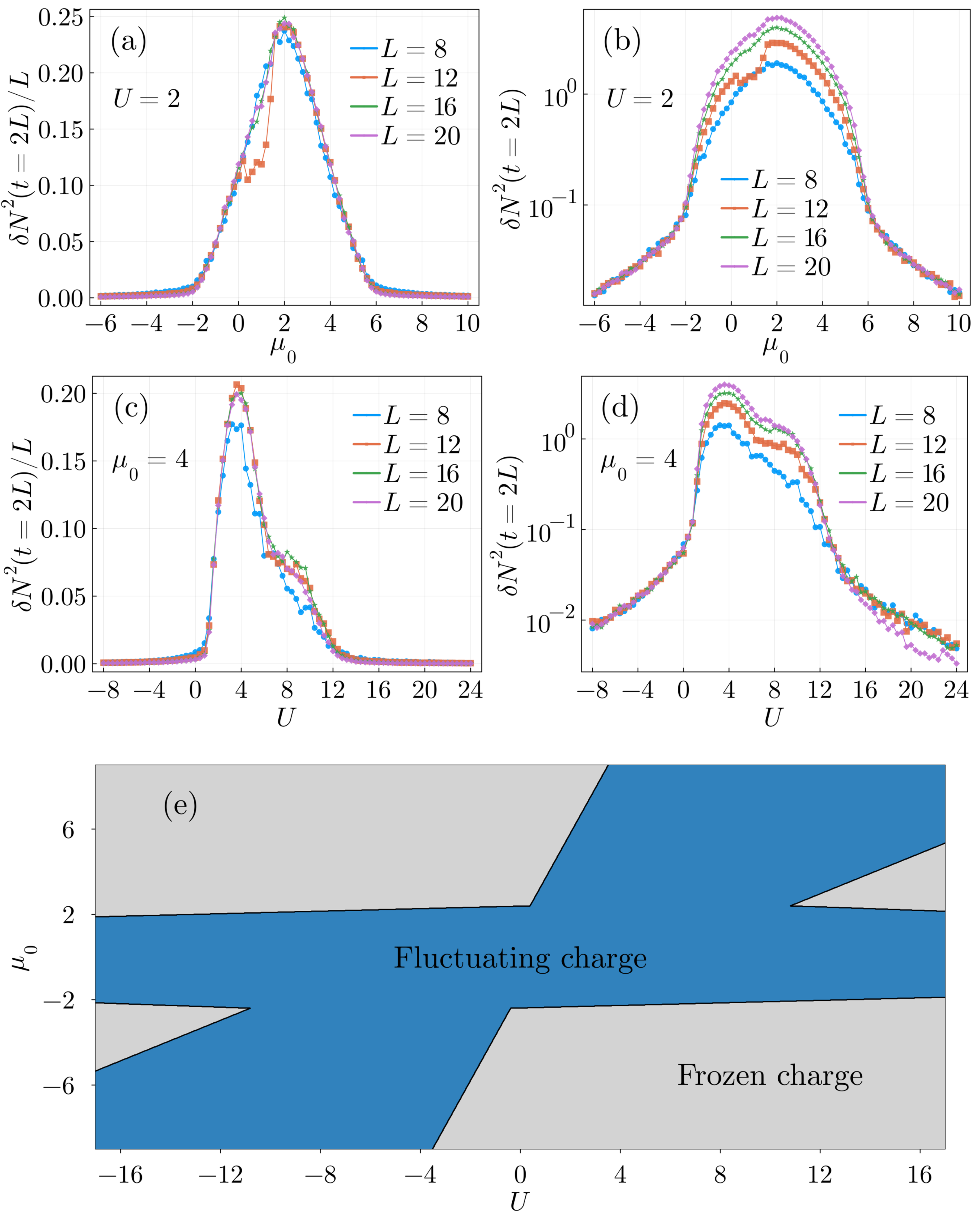}
    \caption{Interacting fermionic chain. (a) The density of charge fluctuation $\delta N^2/L$ as a function of $\mu_0$ for different system sizes. (b) The charge variance $\delta N^2$ as a function of $\mu_0$ for different system sizes on a log-lin scale. Here, $t_0=\Delta=1$, $U=2$, $\nu=0.5$, and the results are averaged over $200$ samples.
    [(c) and (d)] Same as (b) and (c), but with $\mu_0=4$ fixed and $U$ varied.
    (e) Two-dimensional phase diagram of $\mu_0$ and $U$. The blue region represents extensive charge fluctuation and is labeled as ``Fluctuating charge''. The gray region represents negligible charge fluctuation and is labeled as ``Frozen charge''.
 }\label{fig2}
\end{figure}

\section{Boundary perturbation on an interacting fermionic chain}\label{sec::3}
To investigate whether the above phenomena persist in an interacting system, we now consider a simple interacting model by adding an interaction term to the original free $\hat{H}_0$. The new $\hat{H}_0$ becomes
\bea\label{eq::int_H0}
    H_0 &=\sum_{j=1}^L t_0(\hat{c}_j^{\dagger}\hat{c}_{j+1} + \hat{c}_{j+1}^{\dagger}\hat{c}_{j})- \sum_{j=1}^L \mu_0\hat{n}_j + \sum_{j=1}^L U\hat{n}_j \hat{n}_{j+1}.
\eea
The boundary term $\hat{H}_B$ remains unchanged. We prepare a randomly filled product state with a filling factor of $\nu = 0.5$ and study the charge fluctuations in the presence of a boundary perturbation. Numerical results for other filling factors are provided in Appendix \ref{app::2}. We fix $U=2$ and compute $\delta N^2(t = 2L)$ as a function of $\mu_0$. Since the system is no longer a free-fermion dynamics, we simulate its dynamics using exact diagonalization for small system sizes. Results for different system sizes $L$ are shown in Fig.~\ref{fig2}(a) and Fig.~\ref{fig2}(b), where a similar phase transition is observed. The charge fluctuation is extensive for $-2 \lesssim \mu_0 \lesssim 6$ and negligible outside this range.

We can also fix $\mu_0=4$ and vary the interaction  strength $U$. As shown in Fig.~\ref{fig2}(c) and Fig.~\ref{fig2}(d), the results for $\mu_0=4$ indicate that the charge fluctuation is extensive for $1 \lesssim U \lesssim 14$ and negligible outside this range. We also extract a two-dimensional phase diagram by varying $\mu_0$ and $U$, as shown in Fig.~\ref{fig2}(e). It is important to note that the phase boundary may not be entirely accurate and is likely subject to strong finite-size effects due to system size limitations. However, at $U=0$, the phase boundary is approximately consistent with results obtained from larger systems in the previous section.

\begin{figure}
    \centering
    \includegraphics[width=0.99\linewidth]{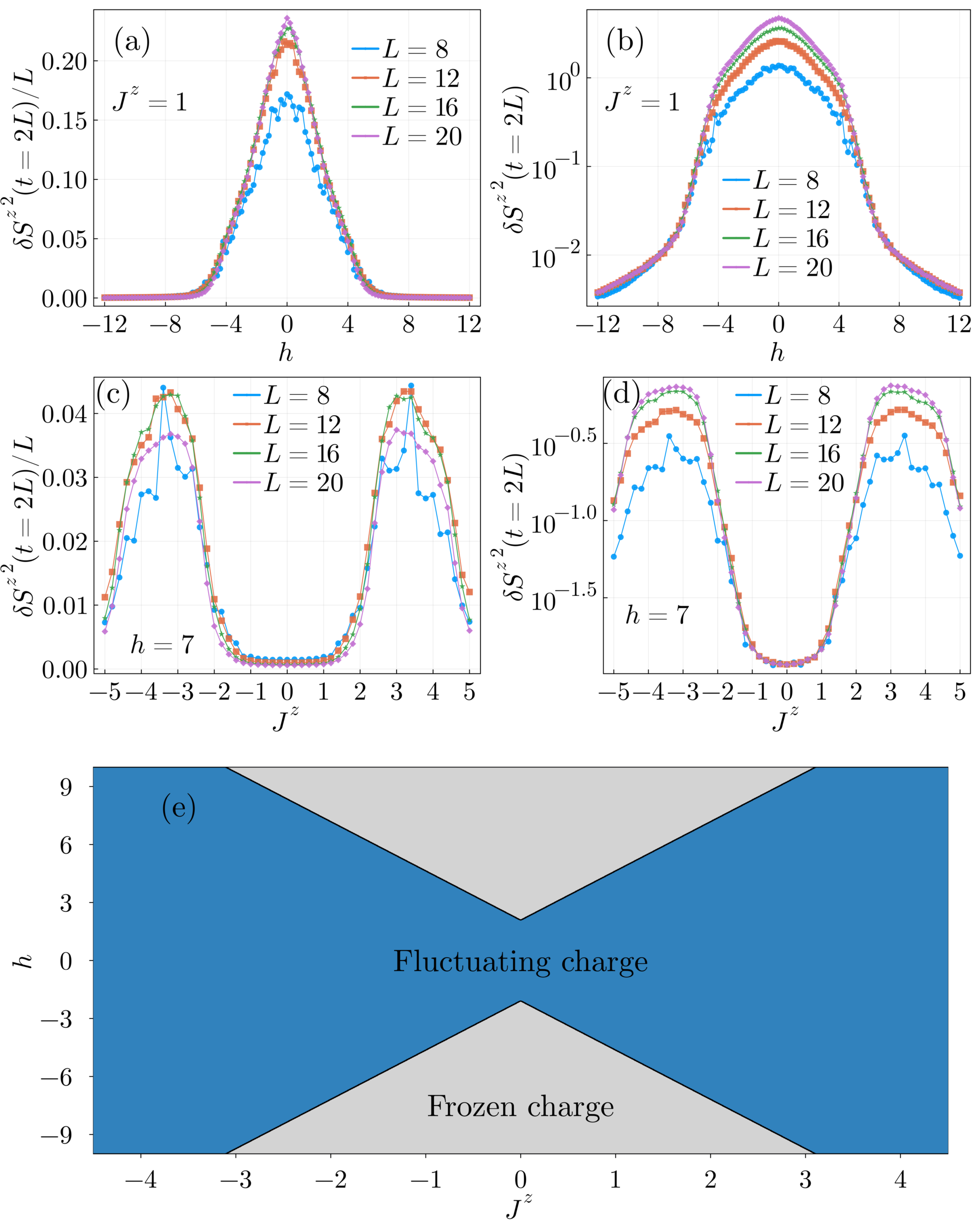}
    \caption{Interacting spin chain. (a) The density of charge fluctuation $\delta {S^z}^2/L$ as a function of $h$ for different system sizes. (b) The charge variance $\delta {S^z}^2$ as a function of $h$ for different system sizes on a log-lin scale. Here, $J^\perp=\Delta=1$, $J^z=1$, and the results are averaged over $200$ samples.
    [(c) and (d)] Same as (b) and (c), but with $h=7$ fixed and $J^z$ varied.
    (e) Two-dimensional phase diagram of $h$ and $J^z$. The blue region represents extensive charge fluctuation and is labeled as ``Fluctuating charge''. The gray region represents negligible charge fluctuation and is labeled as ``Frozen charge''.
 }\label{fig3}
\end{figure}

\section{Boundary perturbation on an interacting spin chain}\label{sec::4}
We consider an interacting spin chain of finite size $L$ with PBC. The bulk Hamiltonian is
\bea
    \hat{H}_0 = \sum_{j=1}^L \Big(J^\perp (\hat{\sigma}_j^x\hat{\sigma}_{j+1}^x+\hat{\sigma}_j^y\hat{\sigma}_{j+1}^y)+J^z\hat{\sigma}_j^z\hat{\sigma}_{j+1}^z\Big) + \sum_{j=1}^L h \hat{\sigma}_j^z,
\eea
and the boundary perturbation is
\bea
    \hat{H}_B = \Delta \hat{\sigma}_1^x.
\eea
Here, $\hat{\sigma}_j^\alpha$ denote Pauli matrices at different lattice sites $j$ with $\alpha=x,y,z$. The bulk Hamiltonian $\hat{H}_0$ is the XXZ spin chain in a magnetic field. Note that $\hat{H}_0$ commutes with the spin operator $\hat{S}^z=\frac12\sum_j\hat{\sigma}_j^z$, so that the total $S^z$ is conserved. The boundary term $\hat{H}_B$ explicitly breaks this conservation law. We prepare the initial state $\ket{\Psi(0)}$ of $\hat{S}^z$ with a fixed  $S^z$, i.e., 
\bea
    \hat{S}^z\ket{\Psi(0)} = S^z\ket{\Psi(0)}.
\eea
For times $t>0$, the system evolves unitarily with $\hat{H}=\hat{H}_0+\hat{H}_B$, and we analyze the spin variance in the z-direction:
\bea
    \delta {S^z}^2(t) = \bra{\Psi(t)} (\hat{S}^z-\bra{\Psi(t)} \hat{S}^z \ket{\Psi(t)})^2 \ket{\Psi(t)}.
\eea

In the numerical study, we prepare a randomly filled product state with half the spins up and half the spins down and compute $\delta {S^z}^2(t=2L)$ as a function of $h$. Results for different system sizes $L$ are shown in Fig.~\ref{fig3}(a)-(d). When $J^z=1$ is fixed, the charge fluctuation is extensive for $-5 \lesssim h \lesssim 5$ and negligible outside this range [Fig.~\ref{fig3}(a) and Fig.~\ref{fig3}(b)]. Note that in Fig.~\ref{fig3}(a), the curve for $L=8$ does not collapse with the other three lines, likely due to strong finite-size effects. When $h=7$ is fixed, the charge fluctuation is negligible for $-2 \lesssim J^z \lesssim 2$ and extensive outside this range [Fig.~\ref{fig3}(c) and Fig.~\ref{fig3}(d)]. Based on the trend of the curve, it is likely that the system will reenter the charge-frozen phase when $|J^z|>5$. However, due to system size limitations, precisely determining the phases in this range is challenging, so we do not present results for this region. We also extract a two-dimensional phase diagram by varying $h$ and $J^z$, as shown in Fig.~\ref{fig3}(e).

\begin{figure}
    \centering
    \includegraphics[width=0.99\linewidth]{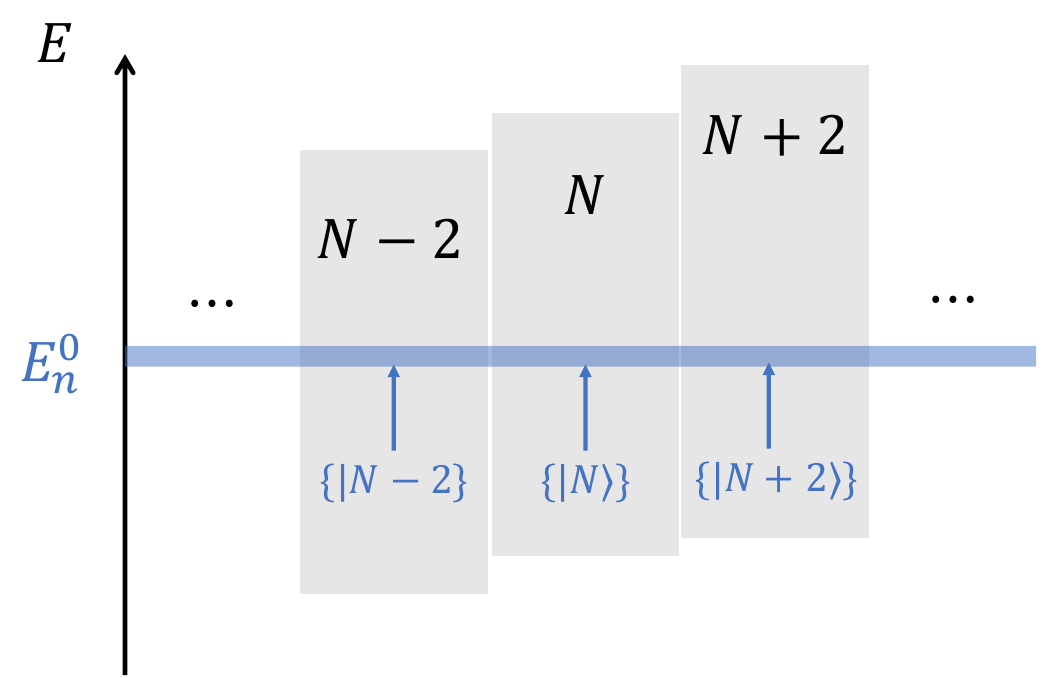}
    \caption{Part of the many-body spectrum of $\hat{H}_0$ of free-fermion chain. The shaded gray area represents eigenstates that share the same labeled particle number: $N-2$, $N$, or $N+2$. $E_n^0$ is one particular eigenvalue of $\hat{H}_0$.
 }\label{spectrum}
\end{figure}

\begin{figure}
    \centering
    \includegraphics[width=0.99\linewidth]{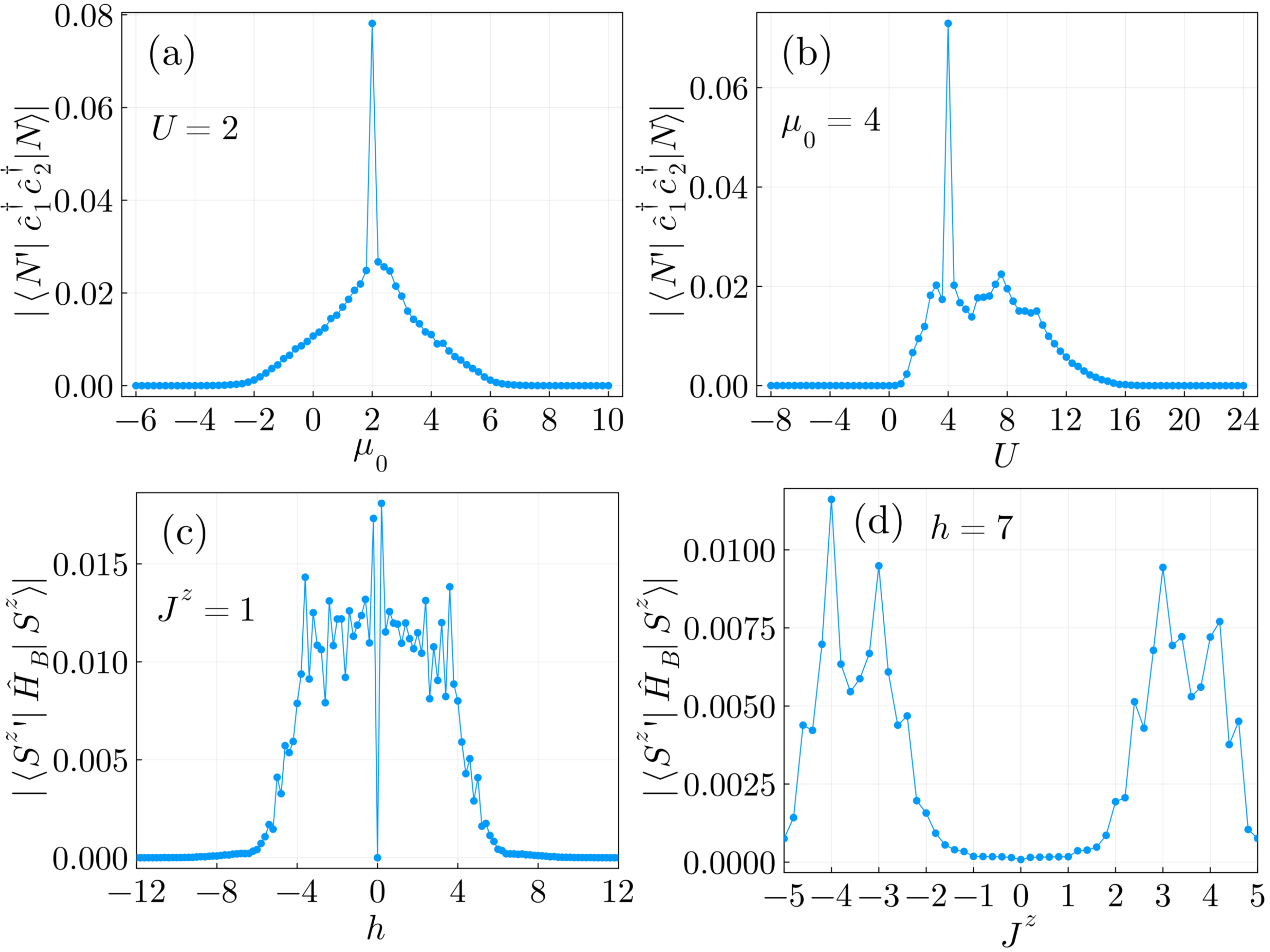}
    \caption{The structure of $\hat{P}\hat{H}\hat{P}$. Interacting fermionic chain: (a) The absolute value of the overlap between $\hat{c}_1^{\dagger}\hat{c}_2^{\dagger}\ket{N}$ and $\ket{N'}$ for different values of $\mu_0$. Here, $L=12$, $t_0=\Delta=1$, $U=2$, and the results are averaged over $1000$ pairs of eigenstates with an energy difference smaller than $0.1$, satisfying $|N'-N-2|<10^{-5}$. [Corresponding to Fig.~\ref{fig2}(a) and Fig.~\ref{fig2}(b)]. (b) Same as (a), but with $\mu_0=4$ fixed and $U$ varied. [Corresponding to Fig.~\ref{fig2}(c) and Fig.~\ref{fig2}(d)]. 
    Interacting spin chain: (c) The absolute value of the overlap between $\hat{H}_B\ket{S^z}$ and $\ket{{S^z}'}$ for different values of $h$. Here, $L=12$, $J^\perp=\Delta=1$, $J^z=1$, and the results are averaged over $1000$ pairs of eigenstates with an energy difference smaller than $0.3$, satisfying $|{S^z}'-S^z-1|<0.005$ or $|{S^z}'-S^z+1|<0.005$. [Corresponding to Fig.~\ref{fig3}(a) and Fig.~\ref{fig3}(b)]. (d) Same as (c), but with $h=7$ fixed and $J^z$ varied. [Corresponding to Fig.~\ref{fig3}(c) and Fig.~\ref{fig3}(d)].
 }\label{fig5}
\end{figure}

\begin{figure}
    \centering
    \includegraphics[width=0.99\linewidth]{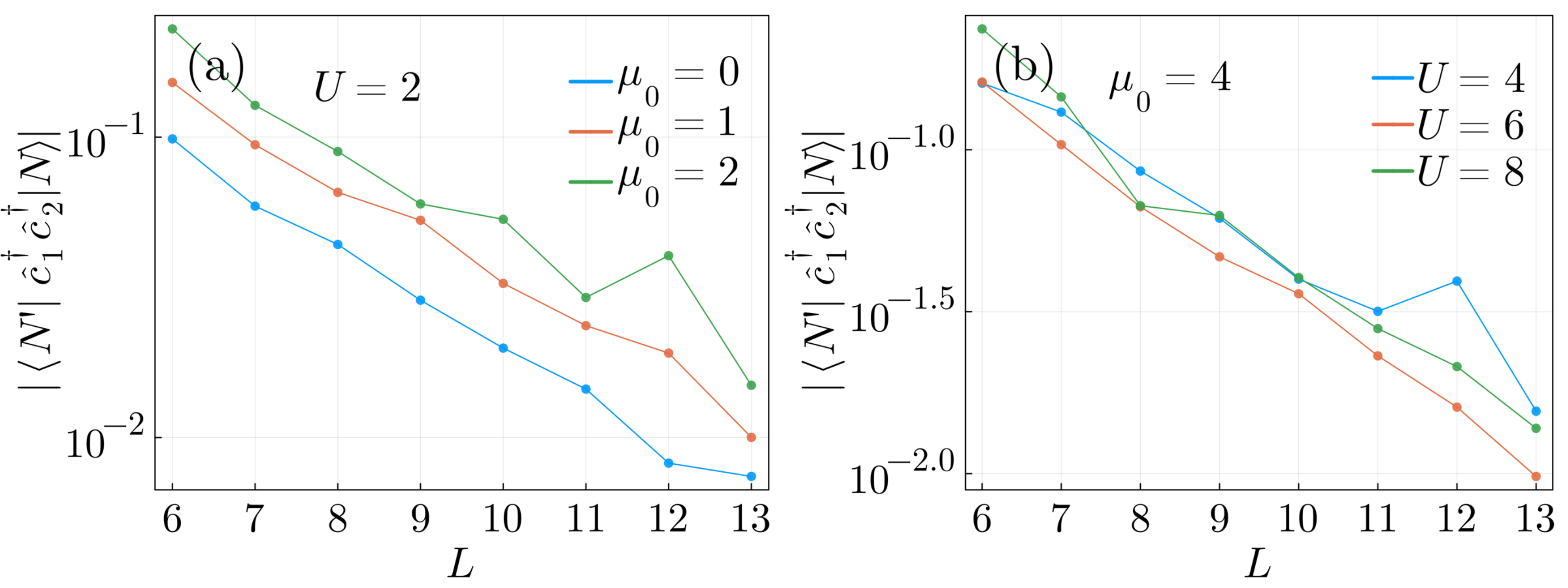}
    \caption{
    The absolute value of the overlap between $\hat{c}_1^{\dagger}\hat{c}_2^{\dagger}\ket{N}$ and $\ket{N'}$ in the charge-fluctuating phase of the interacting fermionic chain for different system sizes. (a) $U=2$ is fixed. (b) $\mu_0=4$ is fixed. Here $t_0=\Delta=1$. For each system size, the data are averaged over 1000 pairs of eigenstates with energy difference smaller than $0.5$ and satisfying $|N'-N-2|<10^{-5}$. The state $\ket{N}$ is chosen from the middle of the spectrum: denoting the maximal and minimal energies by $E_{\mathrm{max}}$ and $E_{\mathrm{min}}$, we require its energy $E$ to satisfy $\big|(E-E_{\mathrm{min}})/(E_{\mathrm{max}}-E_{\mathrm{min}})-0.5\big|<0.3$.
 }\label{r_fig5}
\end{figure}

\section{Effective Hamiltonian}\label{sec::5}
The numerical simulation above suggests that by varying the bulk parameter, the boundary term may induce a bulk phase transition. To better understand this behavior, we employ perturbation theory. Note that the boundary term is supported on a finite number of sites and can only change the total energy by an $O(1)$ constant. Therefore in the perturbation framework, we can restrict our focus to the original eigenstates of $\hat{H}_0$ with a fixed energy $E_0$ and construct the effective Hamiltonian within this subspace. We then analyze the eigenstates of this effective Hamiltonian. 
In Appendix \ref{app::r_1}, we also establish a connection between the static arguments and the dynamical behavior.
Below, we start with the free-fermion chain and then generalize this to the interacting system.

Consider a typical eigenstate $\ket{n^0}$ of $\hat{H}_0$ with energy $E_n^0$ and particle number $N$:
\bea\label{eq::eig_of_h0}
    \ket{n^0} = \hat{c}_{k_1}^\dagger \hat{c}_{k_2}^\dagger\cdots \hat{c}_{k_N}^\dagger\ket{0}, k_1<k_2<\cdots<k_N.
\eea
For a typical $\ket{n^0}$, it is possible to construct other eigenstates with the same energy. These eigenstates can have particle numbers ranging from $N-O(L)$ to $N+O(L)$. We denote these eigenstates as $\{\ket{N_1\equiv N-O(L)}\},\cdots,\{\ket{N}\},\cdots,\{\ket{N_2\equiv N+O(L)}\}$, as illustrated in Fig.~\ref{spectrum}, which provides a schematic representation of the spectrum of $\hat{H}_0$. Here, $\{\ket{M}\}$ indicates a set of states with energy $E_n^0$ and particle number $M$, while $\ket{M}_j$ refers to a specific state within $\{\ket{M}\}$, where the subscript $j$ denotes the $j$-th state.
The boundary term can couple sectors $\{\ket{M}\}$ with the same energy but different particle numbers. In addition, these sectors need to have the same parity. 
Therefore, we focus on sectors with particle numbers that share the same parity as $N$, and we assume that $N_1$ and $N_2$ also have the same parity as $N$. 
To understand the properties of the perturbed state, we employ degenerate perturbation theory and construct the effective Hamiltonian $\hat{H}_{\text{eff}}$ within this subspace at energy $E_0$ \footnote{See, for example, Ref.~\cite{mila2010strong,abrikosov1965quamtum,fetter2012quantum,kempe2006complexity}. Additionally, refer to Appendix \ref{app::3} for the detailed derivation.}
\bea\label{eq::ham_eff}
    \hat{H}_{\text{eff}}(E) = \hat{P}\hat{H}\hat{P} + \hat{P}\hat{H}\hat{Q}\frac{1}{E-\hat{Q}\hat{H}\hat{Q}}\hat{Q}\hat{H}\hat{P},
\eea
where $\hat{P}$ is the projector onto the subspace, 
\bea
    \hat{P}=\sum_{M=N_1}^{N_2}\sum_j \ket{M}_j\bra{M}_j,
\eea
and $\hat{Q}=1-\hat{P}$. Suppose $\hat{H}\ket{\Psi}=E\ket{\Psi}$ in the full Hilbert space. Then, in the subspace projected by $\hat{P}$, we have $\hat{H}_{\text{eff}}(E)(\hat{P}\ket{\Psi})=E(\hat{P}\ket{\Psi})$. 

In the following analysis, we focus on the first term of Eq.~\eqref{eq::ham_eff} and neglect the higher-order corrections from the second term. We find that considering only the first term is sufficient to explain the two distinct phases observed in the numerical simulations above. A detailed justification is provided in Appendix~\ref{app:PHP_criterion}. The first term  $\hat{P}\hat{H}\hat{P}$ has diagonal elements given by $E_n^0$,
\bea
    &\bra{M}\hat{P}\hat{H}\hat{P}\ket{M}\\
    =&\bra{M}\hat{H}\ket{M}\\
    =&\bra{M}(\hat{H}_0+\hat{H}_B)\ket{M}\\
    =&\bra{M}\hat{H}_0\ket{M}=E_n^0.
\eea
Since $\hat{H}_B$ can only change two particles, the nonzero off-diagonal terms occur between sectors where the particle number differs by two. Specifically, consider $\bra{M+2}\hat{P}\hat{H}\hat{P}\ket{M}$,
\bea
    &\bra{M+2}\hat{P}\hat{H}\hat{P}\ket{M}\\
    =&\bra{M+2}\hat{H}\ket{M}\\
    =&\bra{M+2}(\hat{H}_0+\hat{H}_B)\ket{M}\\
    =&\bra{M+2}\hat{H}_B\ket{M}\\
    =&\frac{\Delta}{L}\sum_{k_1k_2}e^{-i(k_1+2k_2)} \bra{M+2}\hat{c}_{k_1}^{\dagger}\hat{c}_{k_2}^{\dagger}\ket{M}.
\eea
When $|\mu_0|>2t_0$, there is a spectral gap for the quasiparticles, and the energies of the states $\hat{c}_{k_1}^{\dagger}\hat{c}_{k_2}^{\dagger}\ket{M}$ and $\ket{M+2}$ can never be equal. Consequently, $H_B$ cannot connect different sectors with the same energy and $\bra{M+2}\hat{P}\hat{H}\hat{P}\ket{M}$ vanishes, indicating that $\hat{P}\hat{H}\hat{P}$ is purely a {\it diagonal} matrix. However, when the gap closes, there are many $(M+2)$–particle states with (almost) the same energy as $\hat c_{k_1}^\dagger \hat c_{k_2}^\dagger\ket{M}$. Each individual matrix element of $\hat H_B$ between $\ket{M}$ and one such state is of order $\Delta/L$, but there are $O(L^2)$ such channels. The small amplitude is therefore compensated by the large number of available final states, and the net hopping strength between the $\{\ket{M}\}$ and $\{\ket{M+2}\}$ sectors remains $O(1)$ in the thermodynamic limit (see Appendix~\ref{app:PHP_criterion} for details).
Therefore, we obtain
\bea\label{eq::mat_php}
    &\hat{P}\hat{H}\hat{P} = \\
    &\begin{pNiceMatrix}[first-col,first-row,small]
     & \{\ket{N_1}\}  & \{\ket{N_1+2}\}&  \{\ket{N_1+4}\}& \ldots & \{\ket{N_2}\}\\
              \{\ket{N_1}\} & E_n^0 & O(1) &  & &\\
              \{\ket{N_1+2}\} & O(1) & E_n^0  & O(1) & & \\
              \{\ket{N_1+4}\} & & O(1) & E_n^0  & \ddots & \\
              \vdots &  &  & \ddots  & \ddots & \ddots \\
              \{\ket{N_2}\} &  &  &   & \ddots  &E_n^0
    \end{pNiceMatrix}.
\eea
Consequently, the eigenstates become superpositions of states with different particle numbers and with extensive charge fluctuations.
This leads to a simple \emph{pumping picture}: the spectral gap of the quasiparticle disfavors the boundary from pumping a particle into the bulk. However, when the gap disappears, the pumping process is not suppressed, and the final state can have a very different particle number.
In Appendix \ref{app::4}, we consider a one-dimensional free-fermion chain divided into two half-chains and study particle transport between them, which can also be analyzed using the simple pumping picture.

We observe that whether $\hat{P}\hat{H}\hat{P}$ is a purely diagonal matrix serves as a criterion for the phase transition. When $\bra{M+2}\hat{P}\hat{H}\hat{P}\ket{M}=0$, the charge cannot be pumped into the system and different charge sectors remain disconnected, indicating the charge-frozen phase. Notably, the second term in Eq.~\eqref{eq::ham_eff} can, in principle, establish connections between sectors of the same energy with particle numbers differing by $O(L)$, which is reflected in the nonzero off-diagonal elements in the effective Hamiltonian. However, this process requires applying $H_B$ many times and is therefore highly suppressed: a path–counting analysis (Appendix~\ref{app:PHP_criterion}) shows that their contribution to the hopping between neighboring charge sectors scales at most as $O(1/L)$ and vanishes in the thermodynamic limit. Consequently, we neglect this term and consider only $\hat{P}\hat{H}\hat{P}$ in our analysis, which is sufficient to capture the essential physics of the transition.

We can examine this criterion in the context of generic models: consider the unperturbed Hamiltonian $\hat{H}_0$ and the charge operator $\hat{O}$, and find eigenstates that diagonalize both $\hat{H}_0$ and $\hat{O}$ with the same energy $E_0$. We denote these eigenstates by their corresponding charges as $\ket{O}$.
We then examine the effective Hamiltonian in the subspace spanned by these eigenstates, defined by the projector $\hat{P}$. In many cases, we believe it is sufficient to consider $\hat{P}\hat{H}\hat{P}$ alone, which we refer to as the generic pumping mechanism. If the effective Hamiltonian allows different charge sectors to connect, the final state can exhibit extensive charge fluctuations. Conversely, if such connections are absent, the final state will exhibit negligible charge fluctuations.
Specifically, for the previous two interacting models, we select an energy $E_0$ and identify all eigenstates of $\hat{H}_0$ with energies close to $E_0$. Among them, we choose a pair of eigenstates, $\ket{O}$ and $\ket{O'}$, whose corresponding charges differ, and compute the matrix element $|\bra{O'}\hat{H}_B\ket{O}|$.
(To illustrate, in the interacting fermionic chain, we choose a pair of eigenstates with the same energy and a particle number difference of two, $\ket{N}$ and $\ket{N+2}$, since the boundary term can only change the particle number by two. We then test how the boundary term connects these two states by calculating the matrix element $|\bra{N+2}\hat{H}_B\ket{N}|$.)
As shown numerically in Fig.~\ref{fig5}, this matrix element undergoes significant variation as the bulk parameter changes. For example, in Fig.~\ref{fig5}(a) for the interacting fermionic chain with $U=2$ fixed, it is nonzero in the range $-2 \lesssim \mu_0 \lesssim 6$, indicating that $\hat{P}\hat{H}\hat{P}$ contains nonzero off-diagonal elements, whereas outside this range it is nearly zero, indicating that $\hat{P}\hat{H}\hat{P}$ can be effectively approximated as a diagonal matrix. These results are consistent with the observed charge fluctuation behavior in Fig.~\ref{fig2}(a) and Fig.~\ref{fig2}(b).

The above result is consistent with the eigenstate thermalization hypothesis (ETH) ~\cite{deutsch1991quantum,srednicki1994chaos,rigol2008thermalization}.
For a generic nonintegrable bulk Hamiltonian $\hat H_0$ with a conserved $U(1)$ charge, ETH is expected to hold separately within each fixed–charge sector.
At a given energy $E$, each fixed–charge
sector then contains $e^{S(E)}$ eigenstates, where $S(E)$ is the thermodynamic
entropy. Although $\hat H_B$ changes the total charge by $\pm 2$ and therefore
connects different charge sectors, in the charge–fluctuating phase our numerics
indicate that the matrix elements connecting eigenstates of similar energy
in neighboring charge sectors obey the standard ETH scaling, $|\bra{m^0}\hat H_B\ket{n^0}|^2 \sim e^{-S(E)}$. In Fig.~\ref{r_fig5}, we calculate $|\bra{N+2}\hat{H}_B\ket{N}|$ for the interacting fermionic chain and find that it decays exponentially with system size. Here, to fix the energy density, we choose $\ket{N}$ to lie in the middle of the spectrum. 
Summing over an exponentially large number of final states thus produces an $O(1)$ hopping amplitude between neighboring charge sectors, so that $\|\hat{P}_{N+2}\hat{H}_B\hat{P}_N\|\sim O(1)$, where $\|\cdot\|$ is the operator norm, in agreement with our free–fermion estimate.
In the charge–frozen regime, by contrast, the blocks $\hat P_{N+2}\hat H_B\hat P_N$ is zero and a similar path counting analysis shows that higher-order terms are strongly suppressed and vanishes in the thermodynamic limit, so the effective dynamics in charge space remains localized.
A more detailed discussion is given in Appendix~\ref{app:PHP_criterion}.

The pumping mechanism is simple yet useful. For example, it offers a natural explanation of the asymmetric profile in Fig.~\ref{fig2}(c), where we fix $\mu_0=4$ and vary the interaction strength $U$, which controls the interaction between neighboring particles. When $U$ is close to $1$, the interaction is weak and it is easier to pump particles into the system, leading to stronger charge fluctuations. As a result, the peak of the curve is skewed toward the lower end of $U$, producing an asymmetric profile.

\begin{figure}
    \centering
    \includegraphics[width=0.99\linewidth]{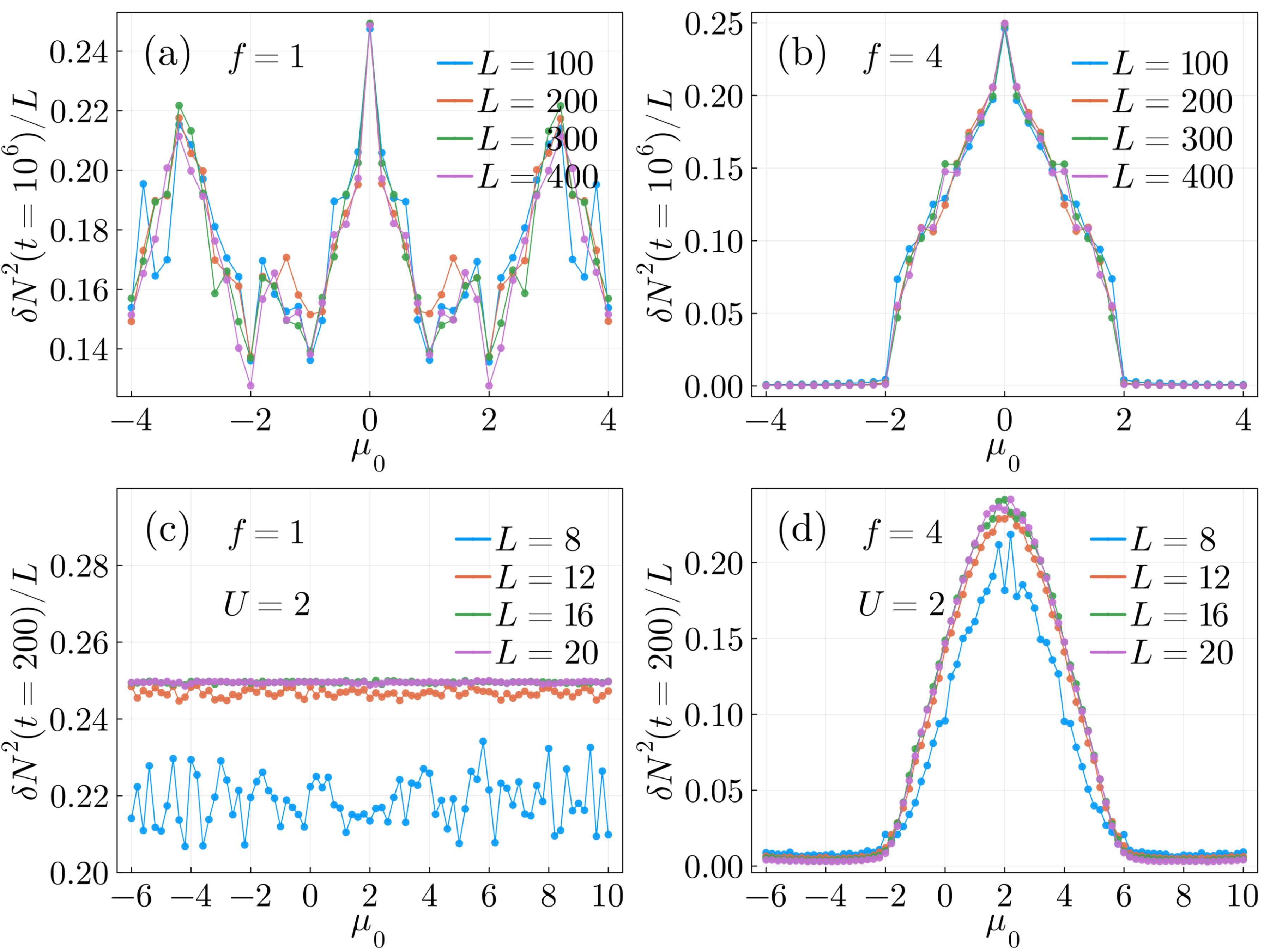}
    \caption{The density of charge variance $\delta N^2/L$ as a function of $\mu_0$ for different system sizes. The system is driven alternately in the bulk and at the boundary according to Eq.~\eqref{eq::floquet}. Free-fermion chain: (a) $f=1$, (b) $f=4$. Interacting fermionic chain with $U=2$: (c) $f=1$, (d) $f=4$. Here, $t_0=\Delta=1$, $\nu=0.5$, and the results are averaged over $100$ samples.
 }\label{r_fig6}
\end{figure}

\begin{figure*}
    \centering
    \includegraphics[width=0.99\linewidth]{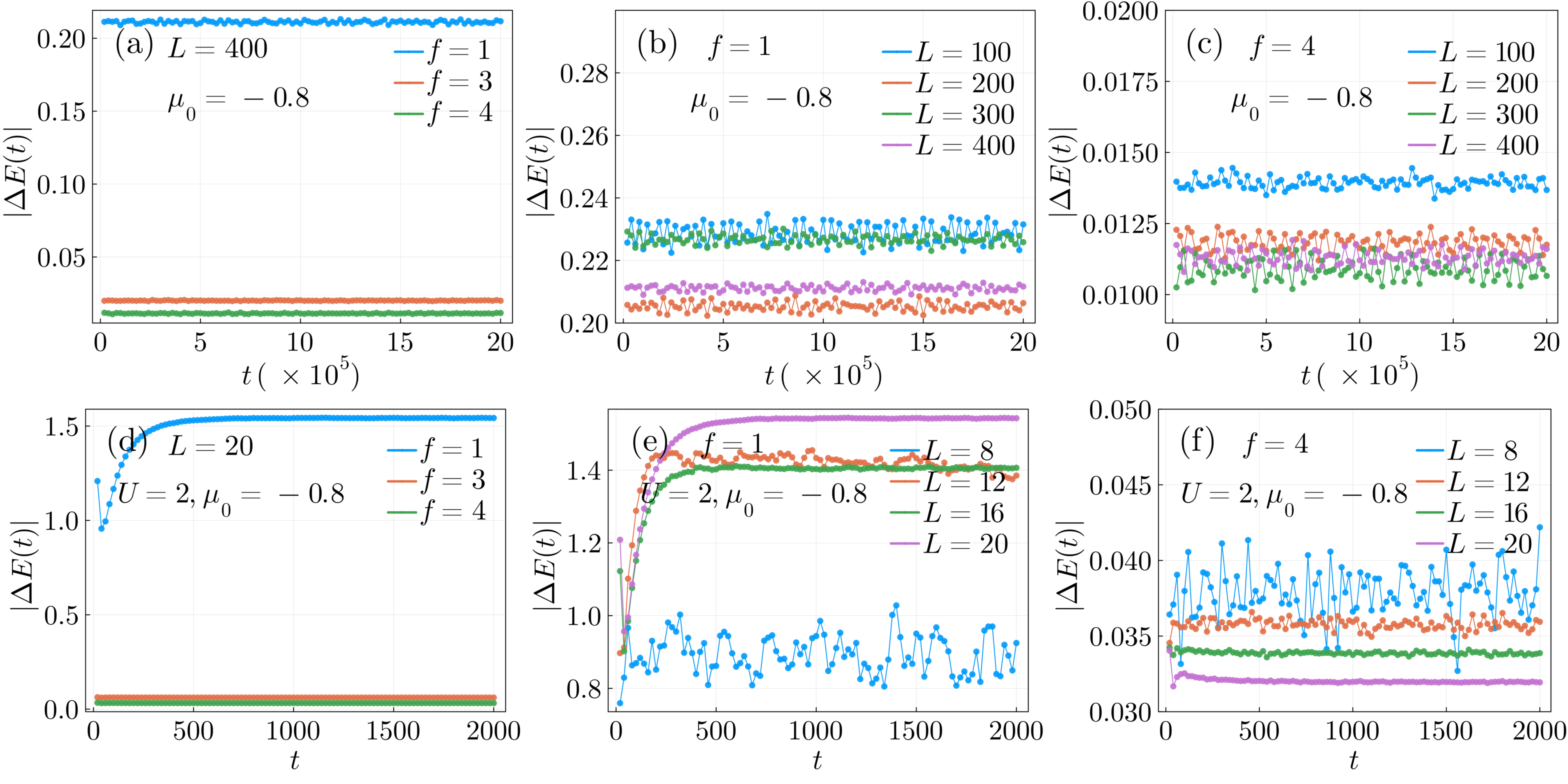}
    \caption{Energy dynamics. The system is driven alternately in the bulk and at the boundary according to Eq.~\eqref{eq::floquet}. Free-fermion chain with $\mu_0=-0.8$: (a) different driving frequencies for $L=400$; (b) different system sizes at $f=1$; (c) different system sizes at $f=4$. Interacting fermionic chain with $U=2, \mu_0=-0.8$: (d) different driving frequencies for $L=20$; (e) different system sizes at $f=1$; (f) different system sizes at $f=4$. Here, $t_0=\Delta=1$, $\nu=0.5$, and the results are averaged over $100$ samples.
 }\label{r_fig7}
\end{figure*}

\begin{figure}
    \centering
    \includegraphics[width=0.6\linewidth]{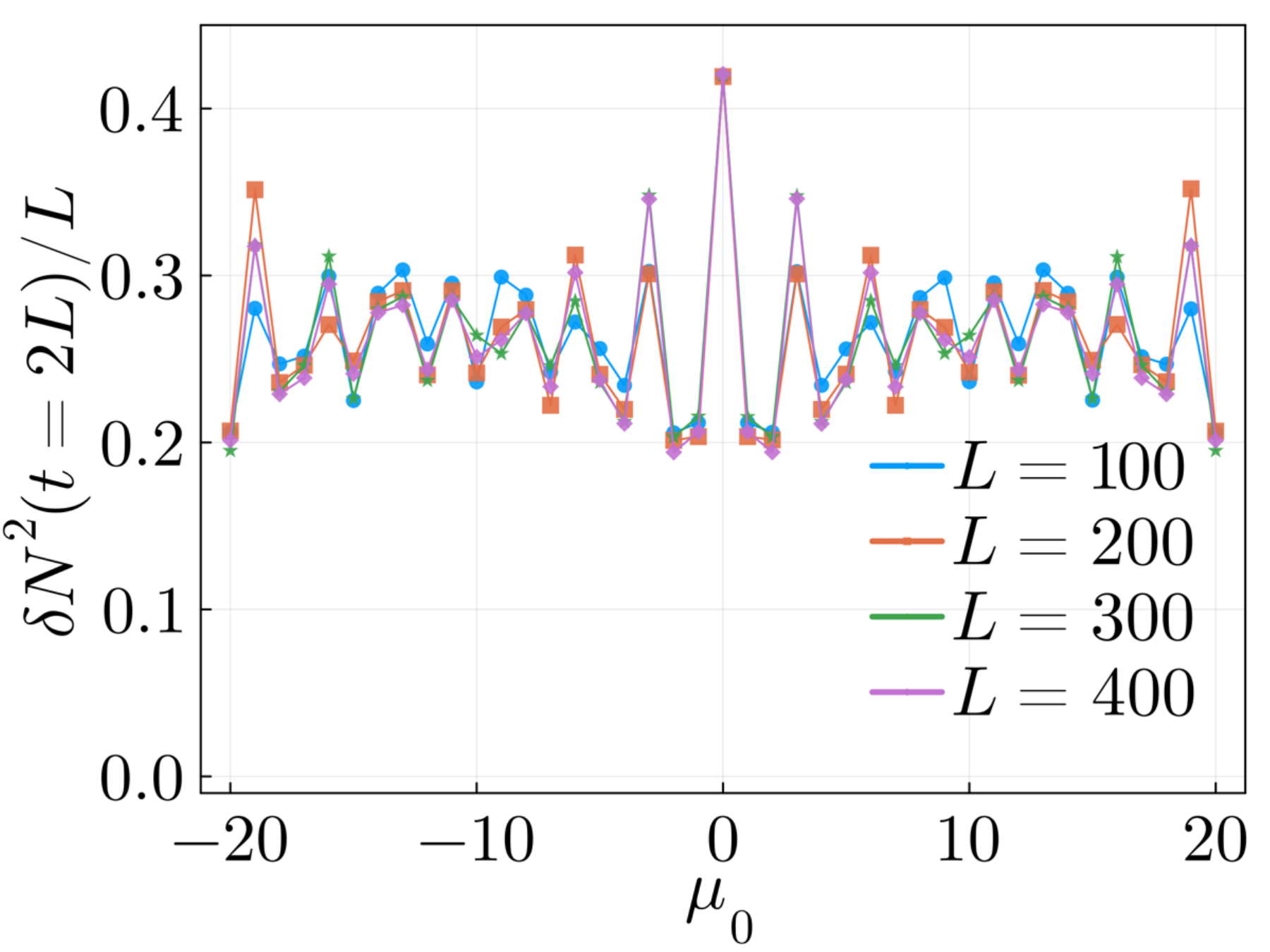}
    \caption{Energy conservation is replaced by conservation of spin in the z-direction in the spinful model. The density of charge variance $\delta N^2/L$ as a function of $\mu_0$ for different system sizes. Here, $t_0=\Delta=1$, and the results are averaged over $1000$ samples.
 }\label{fig6}
\end{figure}

\section{Floquet dynamics}\label{sec::6}
One question that arises is whether the results still hold in Floquet systems, where the system is driven periodically. 
Specifically, we consider a simple ``boundary-kick'' protocol in which the Hamiltonian is time periodic and piecewise constant within each driving period:
\bea
    &\hat H(t)=
    \begin{cases}
    \hat H_0, & 0 \le (t \bmod T) < \tau_0,\\[2pt]
    \hat H_B, & \tau_0 \le (t \bmod T) < T,
    \end{cases} \\
    &\hat H(t+T)=\hat H(t),\quad T=\tau_0+\tau_B,
\eea
where $\hat H_0$ is the bulk Hamiltonian and $\hat H_B$ is the boundary perturbation.
The corresponding stroboscopic evolution over one period is generated by the Floquet operator
\bea\label{eq::U_F}
    U_F \equiv e^{-i\hat{H}_B \tau_B}e^{-i\hat{H}_0 \tau_0}.
\eea
Here we parametrize the protocol by a driving frequency $f$ by choosing $\tau_0=\tau_B=1/f$ (so that $T=2/f$), which gives
\bea
    U_F = e^{-i\hat{H}_B/f}e^{-i\hat{H}_0/f}.
\eea
Starting from an initial state $\ket{\Psi(0)}$, the state after time $t$ is
\bea\label{eq::floquet}
    \ket{\Psi(t)} = U_F^{tf/2}\ket{\Psi(0)},
\eea
where $tf/2$ is the number of Floquet periods elapsed. 
With this protocol, we find that at low driving frequencies the charge-frozen phase disappears.
This occurs because Floquet dynamics generally breaks energy conservation, causing the system to thermalize to infinite temperature in the long-time limit and different charge sectors mix in the presence of boundary perturbation \cite{d2014long,lazarides2014equilibrium}. By contrast, at high driving frequencies, heating can be suppressed for a long time \cite{ho2023quantum,mori2016rigorous,kuwahara2016floquet,abanin2017rigorous,abanin2017effective}. In particular, for a finite system of length $L$, Appendix~\ref{app::r_2} shows that there exists an $L$-dependent threshold $f_c(L)$ above which heating is suppressed at all times even in a generic interacting system. This finite-size high-frequency stabilization is closely related in spirit to the results of Ref.~\cite{d2014long}.
In this regime, the emergence of an effective energy conservation law allows the system to sustain a phase where charge fluctuations remain negligible.

We first consider the free-fermion case, where $\hat{H}_0$ and $\hat{H}_B$ are defined in Eqs.~\eqref{eq::fh0} and \eqref{eq::fhB}. The initial state is a randomly filled product state with a filling factor $\nu$, which is evolved to long times. The steady-state value of $\delta N^2$ is then computed as a function of $\mu_0$. The results for $f=1$ and $f=4$ are shown in Figs. \ref{r_fig6}(a) and \ref{r_fig6}(b). For small $f$ ($f=1$), the charge fluctuation remains extensive, indicating the disappearance of the charge-frozen phase. By contrast, at large $f$ ($f=4$), we observe a phase transition similar to that found under Hamiltonian dynamics.

To explain this phenomenon, we analyze the energy dynamics on such extremely long timescales. Specifically, we consider the energy difference $\Delta E(t)$:
\bea
    \Delta E(t) = \bra{\Psi(t)}\hat{H} \ket{\Psi(t)} - \bra{\Psi(0)} \hat{H} \ket{\Psi(0)},
\eea
where $\hat{H} = \hat{H}_0 + \hat{H}_B$. The results for different frequencies are shown in Figs.~\ref{r_fig7}(a)--(c). The magnitude of $\Delta E(t)$ for large $f$ is much smaller than that for small $f$. In Appendix \ref{app::r_2}, we give a rigorous proof that, for free systems and $f>f_c$, the quantity $\Delta E(t)$ is bounded by a constant independent of the system size.

The above results can be extended to the interacting case. When the interaction term is included, the Hamiltonian $\hat{H}_0$ is modified as defined in Eq.~\eqref{eq::int_H0}. We fix $U=2$ and compute $\delta N^2$ as a function of $\mu_0$. The results for $f=1$ and $f=4$ are shown in Figs. \ref{r_fig6}(c) and \ref{r_fig6}(d). Similar to the free-fermion case, the charge fluctuation remains extensive for small $f$ ($f=1$), while the frozen phase reemerges at large $f$ ($f=4$).  The corresponding energy dynamics are shown in Figs. \ref{r_fig7}(d)--(e): at low frequencies, $\Delta E(t)$ gradually increases and saturates to a finite constant, whereas at high frequencies, $\Delta E(t)$ quickly saturates to a much smaller constant. Here, the charge frozen phase observed at high frequency is a finite size effect. In Appendix~\ref{app::r_2}, we show that this non-heating window shrinks and eventually disappears in the thermodynamic limit~\cite{ho2023quantum,mori2016rigorous,kuwahara2016floquet,abanin2017rigorous,abanin2017effective}.

Another interesting question is whether the phase transition persists when the conservation law is changed.
Here, we replace the conservation of energy with the conservation of spin in the z-direction. Specifically, we consider the non-interacting spinful model: The bulk Hamiltonian is
\bea
    \hat{H}_0 &=\sum_{j,\sigma=\uparrow,\downarrow} t_0(\hat{c}_{j,\sigma}^{\dagger}\hat{c}_{j+1,\sigma} + \hat{c}_{j+1,\sigma}^{\dagger}\hat{c}_{j,\sigma})- \sum_{j,\sigma=\uparrow,\downarrow} \mu_0\hat{n}_{j,\sigma},
\eea
where $\sigma=\uparrow,\downarrow$ is the spin index. The boundary perturbation is
\bea
    \hat{H}_B = \Delta (\hat{c}_{1,\uparrow}^{\dagger}\hat{c}_{2,\downarrow}^{\dagger}+\hat{c}_{2,\downarrow}\hat{c}_{1,\uparrow}).
\eea
Note that both $\hat{H}_0$ and $\hat{H}_B$ commute with the spin operator in the z-direction $\hat{S}^z = \sum_j (\hat{n}_{j,\uparrow}-\hat{n}_{j,\downarrow})/2$~\cite{tasaki2020physics}, whereas only $\hat{H}_0$ commutes with the number operator $\hat{N}=\sum_j(\hat{n}_{j,\uparrow}+\hat{n}_{j,\downarrow})$. We prepare a randomly filled product state with half the spins up and half the spins down and evolve the state periodically with $f=1$ according to Eq.~\eqref{eq::floquet}. At this low driving frequency, the effective energy conservation is no longer valid.
The state is evolved for a long time and the steady-state value of $\delta N^2$ is computed as a function of $\mu_0$. The results for different system sizes are shown in Fig.~\ref{fig6}. The charge fluctuation remains extensive across a wide range of chemical potential values.

\section{Discussion}\label{sec::7}
In this paper, we investigate the impact of boundary perturbations on the dynamics of a Hamiltonian with an additional continuous symmetry. We numerically study both free and interacting models, demonstrate that if the boundary perturbation breaks this continuous symmetry, varying the bulk parameters can give rise to two distinct phases characterized by charge fluctuations: (1) Frozen phase: The charge fluctuation is negligible, and the boundary perturbation does not significantly influence the bulk dynamics. (2) Fluctuating phase: Extensive charge fluctuation occurs, allowing charges to be pumped into or out of the system through the boundary. Using degenerate perturbation theory, we explain this behavior from a simple pumping picture and propose a criterion for determining the phase diagram of a generic model. Furthermore, although our study focuses on one-dimensional models, our findings remain applicable in higher dimensions.

The effective energy conservation is a necessary condition for the existence of the phase transition. We study Floquet dynamics and find that the charge-frozen phase survives at high driving frequencies but disappears at low frequencies. At high frequencies, the frozen phase is strictly stable in the free-fermion limit and in finite-size interacting systems. We further demonstrate that when the effective energy conservation law is broken or replaced by a different conservation law, only a charge-fluctuating phase is observed, suggesting that energy conservation has unique effects.

In the future, it would be intriguing to investigate quantum systems with energy conservation alongside multiple other conservation laws, which may exhibit strong fragmentation behavior~\cite{sala2020ergodicity,moudgalya2022hilbert}. Previous research has shown that introducing boundary perturbations can lead to ultraslow thermalization in such systems~\cite{han2024exponentially}. Exploring the potential transition to charge-frozen phases in these systems would be particularly interesting.

\begin{acknowledgements}
    We thank Vikram Ravindranath, Hanchen Liu, Tianci Zhou, and Yicheng Tang for helpful discussions. Q.G. expresses gratitude to Vikram Ravindranath for his guidance on numerical simulations of Gaussian states. We thank the anonymous referee for valuable comments on the Floquet analysis, which led us to extend our study of the high-frequency regime and the stability of the frozen phase. We gratefully acknowledge computing resources from Research Services at Boston College and the assistance provided by Wei Qiu.
    This research is supported by the National Science Foundation under Grant No. DMR-2219735 (Q.G. and X.C.).
\end{acknowledgements}

\bibliography{manuscript.bbl}

\clearpage
\onecolumngrid
\pagenumbering{arabic} 

\renewcommand{\thepage}{A\arabic{page}}
\renewcommand{\thesection}{A\arabic{section}}
\renewcommand{\theequation}{A\arabic{equation}}
\renewcommand{\thefigure}{A\arabic{figure}}
\renewcommand{\thetable}{A\arabic{table}}

\setcounter{equation}{0}
\setcounter{figure}{0}
\setcounter{table}{0}

\appendix
\section{Particle dynamics and energy dynamics of free-fermion chain}\label{app::1}
\begin{figure}[h]
    \centering
    \includegraphics[width=0.99\linewidth]{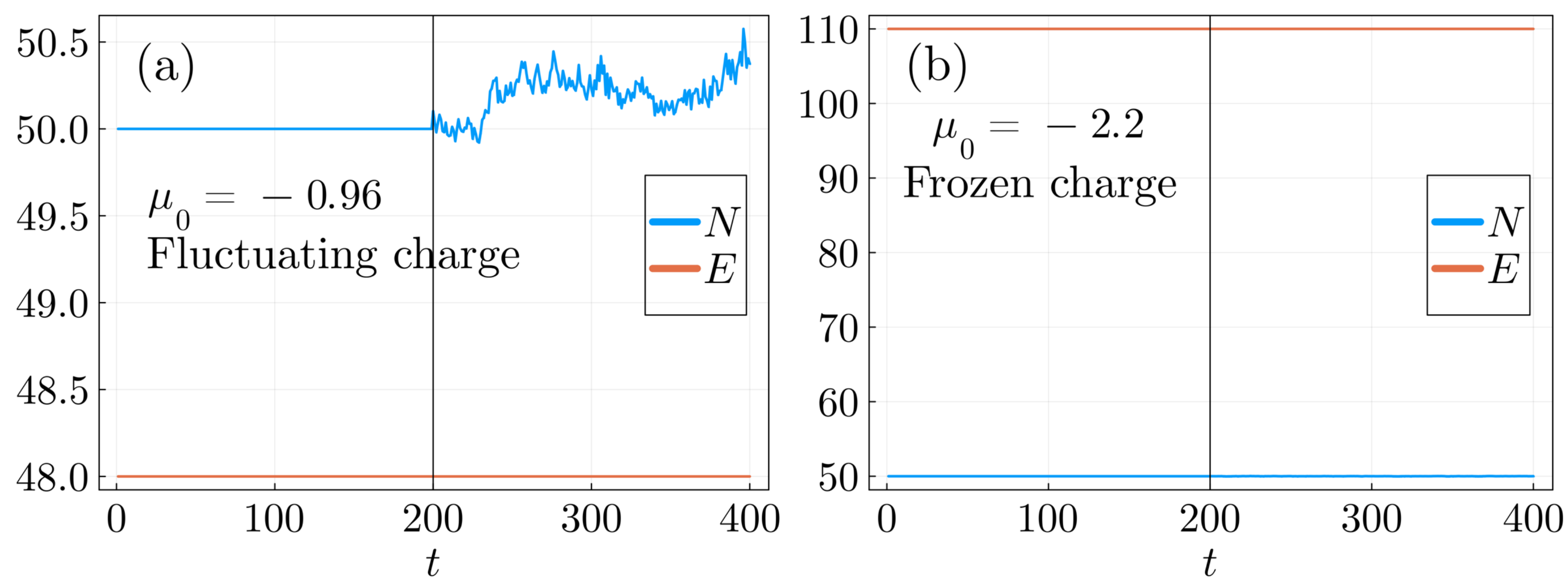}
    \caption{Free-fermion chain. The particle dynamics and energy dynamics for (a) $\mu_0=-0.96$ and (b) $\mu_0=-2.2$. The black vertical lines in both figures indicate the moment when the boundary term is added. Here, $t_0=\Delta=1$, $\nu=0.5$, and $L=100$.
 }\phantomsection\label{sfig1}
\end{figure}

We consider the same free-fermion chain as in the manuscript:
\bea
    \hat{H}_0 =&\sum_{j=1}^L t_0(\hat{c}_j^{\dagger}\hat{c}_{j+1} + \hat{c}_{j+1}^{\dagger}\hat{c}_{j})- \sum_{j=1}^L \mu_0\hat{n}_j,\\
    \hat{H}_B =& \Delta(\hat{c}_1^{\dagger}\hat{c}_2^{\dagger} + \hat{c}_2\hat{c}_1).
\eea
We prepare a randomly chosen initial state $\ket{\Psi(0)}$ with a fixed filling factor $\nu\neq 0$, so that $N=L\nu$:
\bea
    \hat{N}\ket{\Psi(0)} = L\nu\ket{\Psi(0)}.
\eea
To examine how the boundary term $\hat{H}_B$ influences the energy of the system, we first evolve the state unitarily with $\hat{H}_0$ from time $t=0$ to $t=L$. Then, we add the boundary term and continue the unitary evolution with $\hat{H}=\hat{H}_0+\hat{H}_B$ from time $t=L$ to $t=2L$, tracking the energy dynamics throughout. Specifically, the energy of the system from time $t=0$ to $t=L$ is given by $E(t)=\bra{\Psi(t)}\hat{H}_0\ket{\Psi(t)}$, and from time $t=L$ to $t=2L$, the energy is $E(t)=\bra{\Psi(t)}\hat{H}\ket{\Psi(t)}$. We present the particle dynamics and energy dynamics for two examples in different phases in Fig.~\ref{sfig1}. In both phases, the energy remains constant, as the expectation value of the boundary term for a particle-conserving state is zero. However, the particle number undergoes a significant change when the spectral gap disappears, while it remains constant when the spectral gap is present.

Below, we provide a proof that the total energy remains unchanged in both cases. This can be seen by computing the energy difference before and after the switch
\bea
    \Delta E(t) =& \bra{\Psi(0)}e^{i\hat{H}t} \hat{H} e^{-i\hat{H}t}\ket{\Psi(0)} - \bra{\Psi(0)} \hat{H}_0 \ket{\Psi(0)} \\
    =& \bra{\Psi(0)} \hat{H} e^{i\hat{H}t} e^{-i\hat{H}t}\ket{\Psi(0)} - \bra{\Psi(0)} \hat{H}_0 \ket{\Psi(0)}\\
    =& \bra{\Psi(0)} \hat{H} \ket{\Psi(0)} - \bra{\Psi(0)} \hat{H}_0 \ket{\Psi(0)}\\
    =& \bra{\Psi(0)} \hat{H}_B \ket{\Psi(0)}\\
    =& 0,
\eea
where the last equality follows from the fact that the initial state $\ket{\Psi(0)}$ has a fixed particle number, while the boundary perturbation $\hat{H}_B = \Delta(\hat{c}_1^{\dagger}\hat{c}_2^{\dagger} + \hat{c}_2\hat{c}_1)$ changes the particle number. Hence, its expectation value vanishes.

\section{Interacting fermionic chain with a filling factor $\nu=0.25$}\label{app::2}
\begin{figure}
    \centering
    \includegraphics[width=0.99\linewidth]{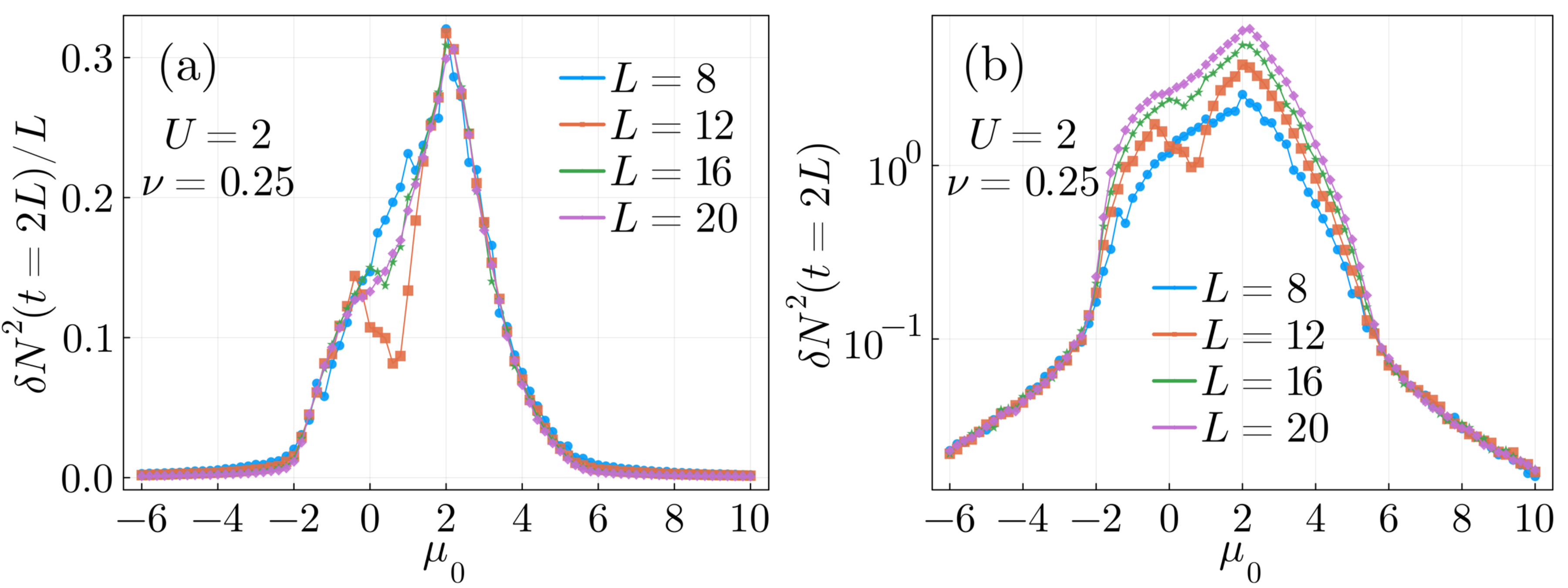}
    \caption{Interacting fermionic chain. (a) The density of charge fluctuation $\delta N^2/L$ as a function of $\mu_0$ for different system sizes. (b) The charge fluctuation $\delta N^2$ as a function of $\mu_0$ for different system sizes on a log-lin scale. Here, $t_0=\Delta=1$, $U=2$, $\nu=0.25$, and the results are averaged over $200$ samples.
 }\phantomsection\label{sfig2}
\end{figure}

In this section, we present the results shown in Fig.~\ref{sfig2} for the interacting model with a filling factor $\nu=0.25$, in comparison to $\nu=0.5$ discussed in the main manuscript. The two results demonstrate that the transition is independent of the filling factor in this model.

\section{Connection between static arguments and dynamical behavior}\label{app::r_1}
In the eigenbasis of $\hat{H}$, we can write the time-evolution operator as
\bea
    e^{-i\hat{H}t} = \sum_n e^{-i E_n t} \ket{n}\bra{n}
\eea
and hence,
\bea
    e^{-i\hat{H}t}\ket{n^0} = \sum_n e^{-iE_n t}\inp{n}{n^0}\ket{n},
\eea
where $\ket{n^0}$ is an eigenstate of the unperturbed Hamiltonian $\hat{H}_0$, and $\ket{n}$ is an eigenstate of the full Hamiltonian $\hat{H}=\hat{H}_0+\hat{H}_B$.

\begin{figure}
    \centering
    \includegraphics[width=0.99\linewidth]{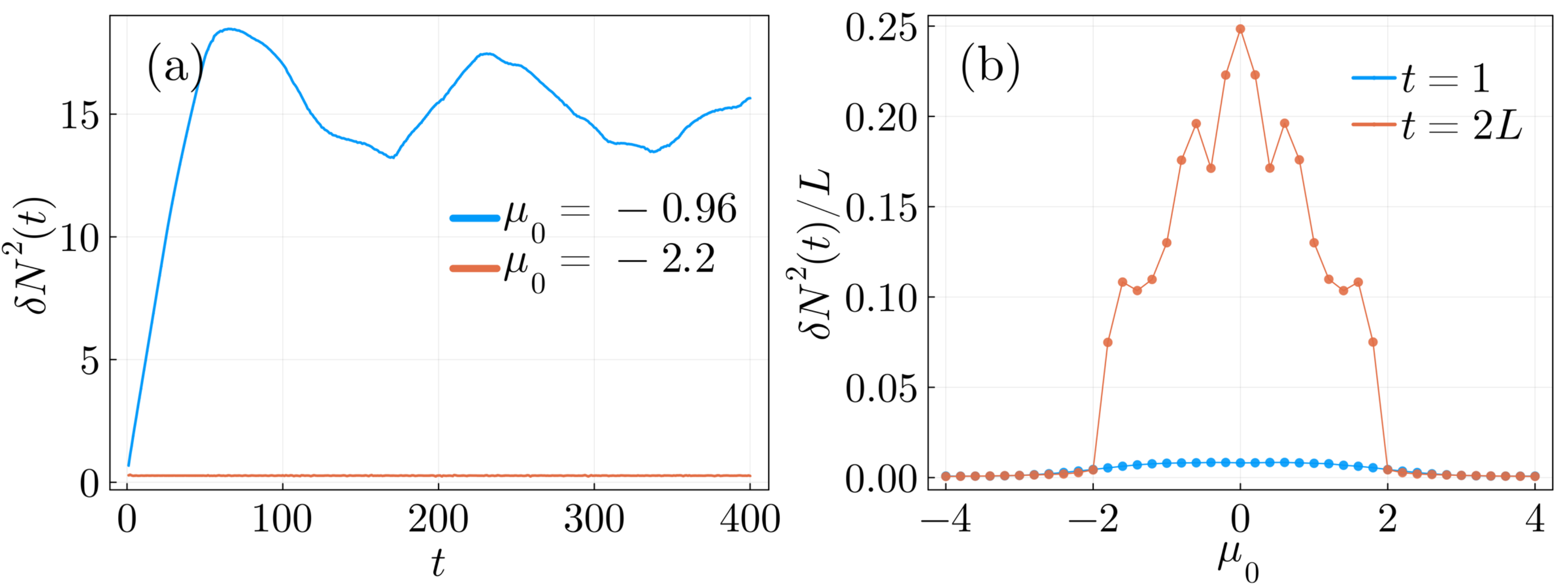}
    \caption{Free-fermion chain. (a) The charge-fluctuation dynamics for $\mu_0=-0.96$ and $\mu_0=-2.2$. (b) The density of charge fluctuation $\delta N^2/L$ as a function of $\mu_0$ at different times. Here, $t_0=\Delta=1$, $\nu=0.5$, $L=100$, and the results are averaged over $1000$ samples.
 }\phantomsection\label{r_a_fig1}
\end{figure}

In the charge-frozen phase, the eigenstates $\ket{n}$ are superpositions of $\ket{n^0}$ with the same particle number $N$, at least when considering the first-order term in degenerate perturbation theory. Since the initial state $\ket{\Psi(0)}$ lies entirely within the sector of fixed particle number $N$, the evolved state at time $t$,
\bea
    \ket{\Psi(t)} = e^{-i\hat{H}t}\ket{\Psi(0)},
\eea
also remains within the same particle-number sector for all $t$. Therefore, charge fluctuations are suppressed at all times, not only in the long-time limit.
(This is neither a prethermal behavior arising from an approximately conserved charge~\cite{ray2020prethermalization}, but a consequence of the charge-frozen eigenstate structure implied by degenerate perturbation theory.)
This conclusion is supported by the numerical data shown in Fig.~\ref{r_a_fig1}(a). From Fig.~\ref{r_a_fig1}(b), we also observe that in order to clearly identify the charge-fluctuating phase, a sufficiently long evolution time is required.

\section{Derivation of the effective Hamiltonian}\label{app::3}
Suppose $\hat{H}\ket{\Psi}=E\ket{\Psi}$, and the full Hilbert space of the system can be decomposed into two orthogonal subspaces, with $\hat{P}$ and $\hat{Q}$ as the corresponding projectors, satisfying $\hat{P}+\hat{Q}=1$. Our goal is to derive the effective Hamiltonian in the subspace projected by $\hat{P}$.

Since $\hat{P}+\hat{Q}=1$, we can rewrite the Schrodinger equation as:
\bea
    \hat{H}(\hat{P}+\hat{Q})\ket{\Psi}=E(\hat{P}+\hat{Q})\ket{\Psi}.
\eea
By projecting separately with $\hat{P}$ and $\hat{Q}$, we obtain:
\begin{align}
\hat{P}\hat{H}\hat{P}\ket{\Psi}+\hat{P}\hat{H}\hat{Q}\ket{\Psi}=&E\hat{P}\ket{\Psi},\label{eq:eff_1}\\
\hat{Q}\hat{H}\hat{P}\ket{\Psi}+\hat{Q}\hat{H}\hat{Q}\ket{\Psi}=&E\hat{Q}\ket{\Psi},\label{eq:eff_2}
\end{align}
and
\bea
    \eqref{eq:eff_2}&\Longrightarrow \hat{Q}\hat{H}\hat{P}\ket{\Psi}=E\hat{Q}\ket{\Psi}-\hat{Q}\hat{H}\hat{Q}\hat{Q}\ket{\Psi}=(E-\hat{Q}\hat{H}\hat{Q})\hat{Q}\ket{\Psi}\\
    &\Longrightarrow \hat{Q}\ket{\Psi} = (E-\hat{Q}\hat{H}\hat{Q})^{-1}\hat{Q}\hat{H}\hat{P}\ket{\Psi},\\
    \eqref{eq:eff_1}&\Longrightarrow \hat{P}\hat{H}\hat{P}\ket{\Psi}+\hat{P}\hat{H}\hat{Q}\hat{Q}\ket{\Psi}=E\hat{P}\ket{\Psi}\\
    &\Longrightarrow \hat{P}\hat{H}\hat{P}\ket{\Psi}+\hat{P}\hat{H}\hat{Q}(E-\hat{Q}\hat{H}\hat{Q})^{-1}\hat{Q}\hat{H}\hat{P}\ket{\Psi}=E\hat{P}\ket{\Psi}\\
    &\Longrightarrow \Big(\hat{P}\hat{H}\hat{P}+\hat{P}\hat{H}\hat{Q}(E-\hat{Q}\hat{H}\hat{Q})^{-1}\hat{Q}\hat{H}\hat{P}\Big)\Big(\hat{P}\ket{\Psi}\Big)=E\Big(\hat{P}\ket{\Psi}\Big).
\eea
The effective Hamiltonian in the subspace projected by $\hat{P}$ is
\bea
    \hat{H}_{\text{eff}}(E)=\hat{P}\hat{H}\hat{P}+\hat{P}\hat{H}\hat{Q}(E-\hat{Q}\hat{H}\hat{Q})^{-1}\hat{Q}\hat{H}\hat{P}.
    \label{Eq:H_eff}
\eea

\section{Why $\hat P \hat H \hat P$ alone suffices as a criterion}
\label{app:PHP_criterion}

In this appendix we explain why it is sufficient to only inspect
\begin{equation}
    \hat P \hat H \hat P
    \;=\;
    \hat P \hat H_0 \hat P
    \;+\;
    \hat P \hat H_B \hat P,
\end{equation}
to distinguish the charge–fluctuating and charge–frozen phases.

Here $\hat H = \hat H_0 + \hat H_B$, where $\hat H_0$ conserves the total particle number $N$, while the boundary pump $\hat H_B$ changes $N$ by $\pm 2$. Consider the eigenstates of $\hat H_0$ with energy $E_0$ within the subspace $\hat P$. In the presence of the boundary perturbation, the total energy can change only by an $O(1)$ amount. Physically, the boundary pump can reshuffle energy only within an $O(1)$ window around $E_0$; it does not modify the energy density in the thermodynamic limit. Therefore, states that originally lie in the subspace $\hat P$ with the same energy $E_0$ broaden into a narrow energy window,
\[
E \in [E_0 - \delta E,\, E_0 + \delta E],
\]
where $\delta E$ is an $O(1)$ constant.

\subsection{Free fermion case}

We first discuss the case where $\hat H_0$ is a free–fermion Hamiltonian.  
Then its many–body eigenstates are Slater determinants, i.e.\ Fock states built from a set of single–particle orbitals (see Eq.~\eqref{eq::eig_of_h0}).
Let $\ket{i}\in \hat P_N$ and $\ket{f}\in \hat P_{N+2}$ be two normalized eigenstates of $\hat H_0$ in the energy window, with particle numbers $N$ and $N+2$, respectively.

In this basis, the boundary term $\hat H_B$ can be written in momentum space as
\begin{align}
    \hat H_B
    &= \frac{\Delta}{L}
       \sum_{k_1,k_2}
       \left(
         e^{-i(k_1+2k_2)}\,\hat c_{k_1}^\dagger \hat c_{k_2}^\dagger
         + e^{i(k_1+2k_2)}\,\hat c_{k_2} \hat c_{k_1}
       \right),
       \label{eq:HB_kspace_appendix}
\end{align}
so that each creation/annihilation of a quasiparticle pair carries an explicit factor $\Delta/L$.  
A standard estimate shows that for any fixed pair of Slater determinants,
\begin{equation}
    \big|\bra{f}\hat H_B\ket{i}\big|
    \;\sim\; \frac{\Delta}{L}.
\end{equation}
However, the window $\hat P_{N+2}$ contains on the order of $\binom{L}{2}\sim L^2$ such states that can be reached by adding two quasiparticles.  
For a fixed normalized $\ket{i}$, the total outgoing weight into the $N+2$ sector is
\begin{equation}
    \big\|\hat P_{N+2}\hat H_B \ket{i}\big\|^2
    \;=\; \sum_{f\in \hat P_{N+2}} |\bra{f}\hat H_B\ket{i}|^2
    \;\sim\; L^2\left(\frac{\Delta}{L}\right)^2
    \;\sim\; O(\Delta^2).
\end{equation}
Hence the operator norm
\begin{equation}
    \big\|\hat P_{N+2}\hat H_B \hat P_N\big\|
    \;\sim\; O(1)
\end{equation}
in the thermodynamic limit.  

This is the situation we refer to as the {\it charge–fluctuating phase}: the in–window block
\begin{equation}
    \hat P_{N+2}\,\hat H\,\hat P_N
    = \hat P_{N+2}\,\hat H_B\,\hat P_N
\end{equation}
already provides an $O(1)$ nearest–neighbor hopping on the ``charge lattice'' labeled by $N$.  
An effective tight–binding model with such couplings naturally supports extended eigenstates in $N$–space and extensive charge fluctuations.

\subsubsection{Charge–frozen phase: organizing virtual processes}

In the {\it charge–frozen phase}, the situation is qualitatively different.  
We propose the following criterion: we say the system is in the charge–frozen phase when, in the thermodynamic limit,
\begin{equation}
    \|\hat P_{N+2}\,\hat H\,\hat P_N\|
    = \|\hat P_{N+2}\,\hat H_B\,\hat P_N\|
    \;\to\; 0,
    \label{eq:PHBP_zero_appendix}
\end{equation}
within the chosen energy window.  
In this regime the processes generated by the second term in Eq.~\eqref{Eq:H_eff} (in Appendix~\ref{app::3}) that try to mix different charge sectors within the window must necessarily leave the $\hat{P}$ space and come back; they are ``virtual'' in the sense that they go through $\hat Q$.

A convenient way to make this structure explicit is to expand the effective Hamiltonian in powers of $\hat H_B$, using the resolvent of $\hat H_0$:
\begin{equation}
    \hat R_0(E)
    := (E - \hat Q \hat H_0 \hat Q)^{-1}.
\end{equation}
This yields a series
\begin{equation}
    \hat H_{\mathrm{eff}}(E)
    \;=\;
    \hat{\mathcal O}^{(1)}(E)
    \;+\;
    \hat{\mathcal O}^{(2)}(E)
    \;+\;
    \hat{\mathcal O}^{(3)}(E)
    \;+\;\cdots,
\end{equation}
with
\begin{align}
    \hat{\mathcal O}^{(1)}(E)
    &= \hat P \hat H_0 \hat P
    \;+\;
    \hat P \hat H_B \hat P,
    \label{eq:O1_def_appendix_again}\\[4pt]
    \hat{\mathcal O}^{(2)}(E)
    &= \hat P \hat H_B \hat Q \,\hat R_0(E)\,\hat Q \hat H_B \hat P,
    \label{eq:O2_def_appendix_again}\\[4pt]
    \hat{\mathcal O}^{(m)}(E)
    &= \hat P \hat H_B \hat Q\hat R_0(E)\,
       \big(\hat Q \hat H_B \hat Q \hat R_0(E)\big)^{m-2}
       \hat Q \hat H_B \hat P,
       \qquad m\ge 3.
    \label{eq:Ok_def_appendix_again}
\end{align}
The term $\hat{\mathcal O}^{(m)}(E)$ contains $m$ insertions of the boundary pump $\hat H_B$ and $(m-1)$ insertions of the resolvent $\hat R_0(E)$.

In the charge–fluctuating phase, $\hat P_{N+2}\hat{\mathcal O}^{(1)}(E)\hat P_N = \hat P_{N+2}\hat H_B\hat P_N$ is already $O(1)$ and dominates.  
In the charge–frozen phase, \eqref{eq:PHBP_zero_appendix} implies
\begin{equation}
    \hat P_{N+2}\hat{\mathcal O}^{(1)}(E)\hat P_N
    = \hat P_{N+2}\hat H_B\hat P_N
    \approx 0.
\end{equation}  
We now estimate the higher order terms $\hat{\mathcal O}^{(m\ge 2)}(E)$.

\subsubsection{Path counting with an energy constraint}\label{subsec::path_counting}

Consider a matrix element
\begin{equation}
    \bra{f}\hat{\mathcal O}^{(m)}(E)\ket{i},
\end{equation}
where $\ket{i}\in\hat P_N$ and $\ket{f}\in\hat P_{N+2}$ are normalized shell states.  
Each insertion of $\hat H_B$ in \eqref{eq:HB_kspace_appendix} creates or annihilates a pair of quasiparticles with momenta $(k_1,k_2)$, and carries a factor $\Delta/L$.  
A given $m$th–order process therefore corresponds to specifying $2m$ momentum indices
\begin{equation}
    (k_1,k_2,\ldots,k_{2m}),
\end{equation}
one pair for each $\hat H_B$.  

\begin{figure}
    \centering
    \includegraphics[width=0.99\linewidth]{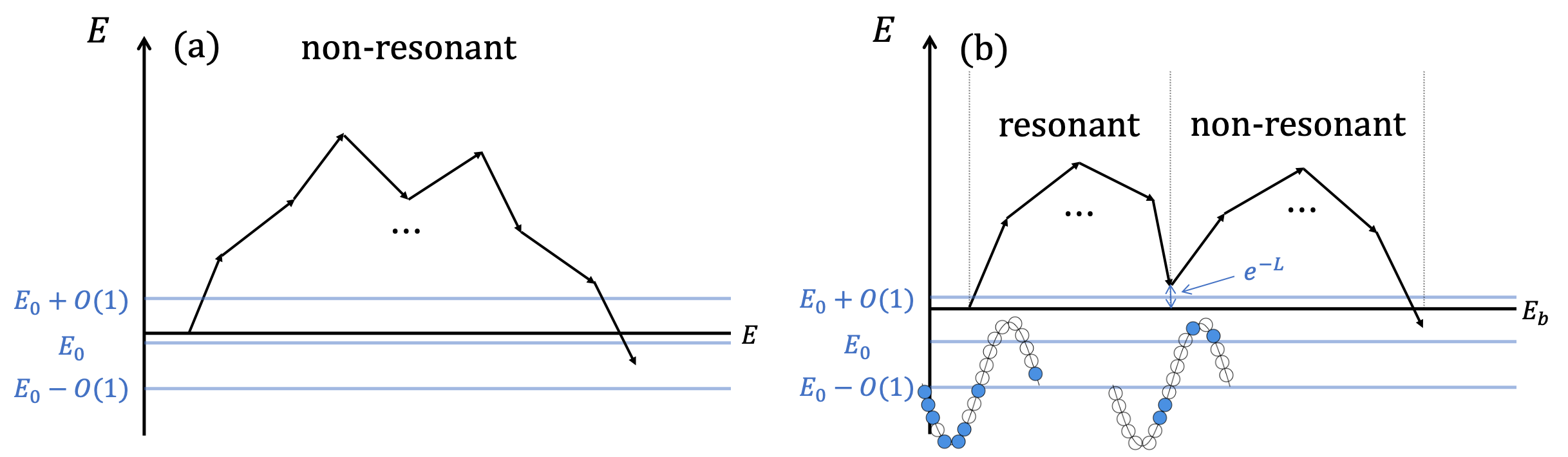}
    \caption{Paths in the charge-frozen phase. (a) Non-resonant paths, in which all intermediate denominators satisfy $|E - E_{\alpha_\ell}^{(0)}|\geq O(1)$. (b) Paths consist of both non-resonant segments and resonant segments. The energy difference between the beginning and end of a resonant part can be exponentially small in $L$, and the corresponding nearly degenerate eigenstates typically differ in the occupation of $O(L)$ quasiparticles.
 }\phantomsection\label{fig:PHP}
\end{figure}

If we ignore the energy window completely, each $k_j$ can independently take $L$ values, so the total number of such ``paths'' is
\begin{equation}
    N_{\rm paths}^{\rm (naive)}(m)
    \;\sim\; L^{2m}.
\end{equation}
For any fixed choice of these indices, the amplitude of the corresponding path scales as
\begin{equation}
    \text{(single-path amplitude)}
    \;\sim\;
    \left(\frac{\Delta}{L}\right)^m \times
    \prod_{\ell=1}^{m-1}\frac{1}{E - E_{\alpha_\ell}^{(0)}},
\end{equation}
where $\{\alpha_\ell\}$ are intermediate many–body states living in the subspace $\hat{Q}$. For a generic path, it can include both non-resonant and resonant segments. The non-resonant part satisfies
\begin{equation}
    |E - E_{\alpha_\ell}^{(0)}|
    \;\geq\; O(1).
\end{equation}
Resonances arise from accidental degeneracies in which two states—one from $\hat{P}$ and one from $\hat{Q}$ become nearly degenerate. As illustrated in Fig.~\ref{fig:PHP}, the energy difference between the beginning and end of such a resonant segment can be exponentially small,
\begin{equation}
    |E_{\mathrm b} - E_\alpha^{(0)}|
    \;\sim\; e^{-cL},
\end{equation}
of the same order as the many-body gap of the system.
In generic situations, these nearly degenerate eigenstates differ in the occupations of
$n_\text{dif}=O(L)$ quasiparticles. Connecting them by repeated
applications of the local boundary term, which creates or annihilates only
two quasiparticles at a time, requires parts of length
$m_\text{r}\sim n_\text{dif}$: schematically, one has to annihilate all
quasiparticles present in one state and recreate those of the other.  The
number of such long paths grows like $n_\text{dif}!$, while each path carries
a factor $(\Delta/L)^{m_\text{r}}$ with $m_\text{r}=O(L)$. Using Stirling’s
approximation, this gives a total weight of order
${n_\text{dif}!}/{L^{m_\text{r}}} \sim L!/L^L \sim e^{-L}$, i.e.\ exponentially
small in $L$ up to polynomial prefactors. This exponentially small weight, when divided by the exponentially small gap, could contribute only an $O(1)$ factor and prevent the resolvent from diverging. Consequently, resonant processes can at most produce a slight renormalization of the phase boundary and do not affect the stability of the charge-frozen phase.
For the scaling estimates below we may therefore safely consider only the
non–resonant parts and bound all denominators by an $L$–independent constant
of order $O(1)$.

If all $L^{2m}$ paths could start from the window and end in the window, the total probability for a charge transfer at order $m$ would be of order
\begin{equation}
    \text{(total probability)}^{\rm (naive)} =
    N_{\rm paths}^{\rm (naive)}\times
    \big|\text{single-path amplitude}\big|^2
    \sim L^{2m}\left(\frac{\Delta}{L}\right)^{2m} \sim O(1),
\end{equation}
i.e.\ virtual processes would produce an $O(1)$ effective hopping, as in a tight–binding chain.

The key point is that the energy window drastically reduces the number of allowed paths.  
In order for a path to contribute to $\bra{f}\hat{\mathcal O}^{(m)}\ket{i}$, it must satisfy:

\begin{itemize}
    \item[(i)] The initial state $\ket{i}$ lies in the window $[E-\delta E,E+\delta E]$.
    \item[(ii)] After $m$ applications of $\hat H_B$, the final state $\ket{f}$ also lies in the same window and has particle number $N+2$.
    \item[(iii)] All intermediate states in the resolvents lie outside the window (in $\hat Q$) and hence differ in energy from $E$ by at least $\delta E$.
\end{itemize}

Condition (ii) is an ``energy–sum'' constraint: the total change in the unperturbed energy induced by the $m$ pair–creation / pair–annihilation events must bring the system back into a narrow window of width $O(1)$ around $E$.  
In a free–fermion picture, this can be written schematically as
\begin{equation}
    \sum_{\text{created }k} \varepsilon(k)
    \;-\;
    \sum_{\text{annihilated }k} \varepsilon(k)
    \;\in\; [-\delta E,\delta E],
\end{equation}
where $\varepsilon(k)$ is the single–particle dispersion.  
For generic, non–fine–tuned dispersions, imposing one such energy constraint effectively reduces the number of independent momentum indices by one: one can think of freely choosing $2m-1$ of the $k_j$, while the last momentum is then fixed by the requirement that the total energy falls back into the window.  
Thus the number of paths that both leave and re–enter the window scales as
\begin{equation}
    N_{\rm paths}^{\rm (window)}(m)
    \;\sim\; L^{2m-1},
\end{equation}
rather than $L^{2m}$.

Combining this reduced path counting with the single–path amplitude, we obtain the following scaling estimate for the total probability of a charge transfer mediated by $m$ virtual insertions of $\hat H_B$:
\begin{equation}
    \text{(total probability at order $m$)}
    \;\sim\;
    N_{\rm paths}^{\rm (window)}\,
    \big|\text{single-path amplitude}\big|^2
    \;\sim\;
    L^{2m-1}
    \left(\frac{\Delta}{L}\right)^{2m}
    \;\sim\;
    \frac{1}{L}.
\end{equation}
For any fixed $m\ge 2$, this weight vanishes as $1/L$ in the thermodynamic limit.

\subsection{Generic interacting systems}

For a generic interacting $\hat H_0$ there is no simple quasiparticle band
structure, so we appeal to the eigenstate thermalization hypothesis
(ETH)~\cite{deutsch1991quantum,srednicki1994chaos,rigol2008thermalization}.
Strictly speaking, the standard ETH ansatz is formulated for nonintegrable
systems without additional conserved quantities beyond energy; in the
presence of a global $U(1)$ charge it is understood to hold separately within
each fixed–charge sector.
For a local operator $\hat O$ which conserves the charge, the matrix
elements in the energy eigenbasis of $\hat H_0$ take the form within a given
symmetry block~\cite{srednicki1999approach}
\begin{equation}
    \bra{m^0}\hat O\ket{n^0}
    =
    \mathcal O(E)\,\delta_{mn}
    + e^{-S(E)/2}\,f(E,\omega)\,R_{mn},
\end{equation}
where $E=(E_m+E_n)/2$, $\omega=E_m-E_n$, $S(E)$ is the thermodynamic
entropy at energy $E$ (so that the number of eigenstates in an $O(1)$
window around $E$ scales as $e^{S(E)}$), $f(E,\omega)$ and
$\mathcal O(E)$ are smooth functions, and $R_{mn}$ is a random number
of order unity.
In the charge-fluctuating phase, we numerically confirm that the matrix elements of
$\hat H_B$ between states with  small $\omega$ and in charge sectors that differ by two
obey the same ETH scaling (see Fig.~\ref{r_fig5} in the main text).
Below we use this to illustrate why the
on–shell block $\hat P\hat H\hat P$ suffices as a criterion for distinguishing the charge-fluctuating and charge-frozen phases.

Within an $O(1)$ energy window around $E$, the total number of
eigenstates scales as $e^{S(E)}$.  In the charge–fluctuating phase,
the projectors $\hat P_N$ and $\hat P_{N+2}$ each contain an
exponentially large number of such eigenstates, and a normalized state
$\ket{i}\in\hat P_N$ is connected by $\hat H_B$ to a finite fraction of
the states in $\hat P_{N+2}$ within the same window.  For a typical
final state $\ket{f}\in \hat P_{N+2}$, ETH then implies (as we indeed observe numerically in Fig.~\ref{r_fig5})
\begin{equation}
    |\bra{f}\hat H_B\ket{i}|^2 \sim e^{-S(E)}.
\end{equation}
Summing over all such $\ket{f}$ in $\hat P_{N+2}$ gives
\begin{equation}
    \big\|\hat P_{N+2} \hat H_B \ket{i}\big\|^2
    = \sum_{f\in \hat P_{N+2}} |\bra{f}\hat H_B\ket{i}|^2
    \;\sim\; e^{S(E)} \times e^{-S(E)} \sim O(1),
\end{equation}
so that the shell–resolved block
$\|\hat P_{N+2}\hat H\hat P_N\|=\|\hat P_{N+2}\hat H_B\hat P_N\|$
remains of order unity in the thermodynamic limit, in complete analogy
with the free–fermion example.

In the charge–frozen phase, our criterion is that the first–order block
between $N$ and $N+2$ sectors vanishes in the thermodynamic limit,
\begin{equation}
    \|\hat P_{N+2}\hat H_B\hat P_N\| \;\to\; 0.
\end{equation}
In terms of ETH, this means that $f(E,\omega\to 0)\to 0$.
The processes that could change the
charge within the window are then the virtual terms
$\hat{\mathcal O}^{(m\ge 2)}(E)$ in the effective Hamiltonian.  As in the
free–fermion case, these processes must leave the energy window and come
back, so their intermediate states are constrained to lie in $\hat Q$,
and the requirement that both the initial and final states lie in the
same narrow window imposes a nontrivial energy–sum constraint on the
allowed paths.

At order $m$, a virtual process corresponds to a path of length $m$ in
the many–body spectrum: a sequence of intermediate eigenstates of
$\hat H_0$ outside the window, connected by insertions of $\hat H_B$.
For each insertion of $\hat H_B$, there are $e^{S(E)}$ possible intermediate eigenstates, while we expect that each path contributes at most a weight of order $e^{-S(E)}$ in probability. Naively, at fixed order $m$, the total probability of a charge transfer remains $O(1)$.
Imposing the energy constraint (that the path starts and ends inside the same
$O(1)$–wide window) is analogous to the free–fermion case: it effectively
reduces the number of independent intermediate choices and hence the
number of admissible paths.  For any fixed $m$, the exponential
suppression in $L$ coming from the ETH factor dominates over the
exponential growth in $N_{\text{paths}}(m)$, so the total contribution
of $\hat{\mathcal O}^{(m)}(E)$ to the block
$\hat P_{N+2}\hat H_{\text{eff}}(E)\hat P_N$ remains exponentially
small in $L$.

Rare resonant segments correspond to parts of a path for which the
energy difference between their endpoints is exponentially small in $L$.
In the interacting case, rather than in the free case, the exponential
suppression coming from the energy–sum constraint itself can compete with
the potential enhancement due to small denominators.  As a result, the
net contribution of these resonant parts may at most slightly
renormalize the location of the phase boundary, but cannot
destroy the charge–frozen phase.

In this sense, the interacting model has the same
qualitative picture as in the free–fermion case: the on–shell block
$\hat P\hat H\hat P$ suffices to determine whether the system is in the charge-fluctuating or charge-frozen phase, while the virtual
terms $\hat{\mathcal O}^{(m\ge 2)}(E)$ provide only parametrically small
corrections that vanish in the thermodynamic limit (at least in the charge-frozen phase).

\section{Particle transport in a free-fermion chain}\label{app::4}
\begin{figure}
    \centering
    \includegraphics[width=0.99\linewidth]{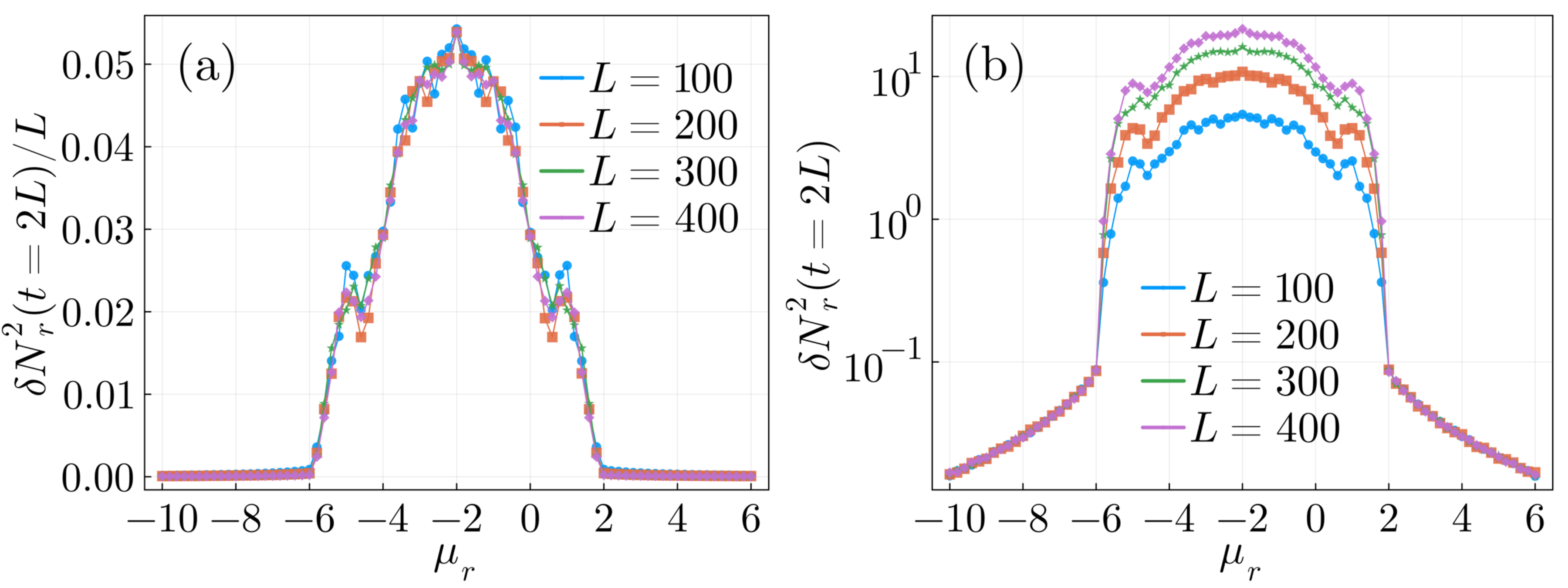}
    \caption{Transport in a free-fermion chain. (a) The density of charge fluctuation $\delta N_r^2/L$ as a function of $\mu_r$ for different system sizes. (b) The charge fluctuation $\delta N_r^2$ as a function of $\mu_r$ for different system sizes on a log-lin scale. Here, $t_l=t_r=\Delta=1$, $\mu_l=-2$, $\nu_l=0.5, \nu_r=0.25$, and the results are averaged over $1000$ samples.
 }\phantomsection\label{sfig3}
\end{figure}

The gap in the bulk spectrum can also influence the transport from other systems. We consider the Hamiltonian of a one-dimensional free-fermion chain of finite size $L$ with open boundary conditions:
\bea
    \hat{H}_0 =& \hat{H}_l+\hat{H}_r,\\
    \hat{H}_l =& \sum_{j\in l} t_l(\hat{c}_j^{\dagger}\hat{c}_{j+1} + \hat{c}_{j+1}^{\dagger}\hat{c}_{j})-\sum_{j\in l} \mu_l\hat{n}_j,\\
    \hat{H}_r = & \sum_{j\in r} t_r(\hat{c}_j^{\dagger}\hat{c}_{j+1} + \hat{c}_{j+1}^{\dagger}\hat{c}_{j})-\sum_{j\in r} \mu_r\hat{n}_j.
\eea
Here, the system is divided into two half-chains, with the subscript $l$ and $r$ denoting the left and right parts. Note that $\hat{H}_0$ commutes with the number operator of the right part, $\hat{N}_r=\sum_{j\in r} \hat{n}_j$. Initially, we fill the left part with a filling factor $\nu_l$ and the right part with a filling factor $\nu_r$. 
We add a coupling term $\hat{H}_C$ between the left and right parts:
\bea
    \hat{H}_C = \Delta(\hat{c}_{L/2}^{\dagger}\hat{c}_{L/2+1} + \hat{c}_{L/2+1}^{\dagger}\hat{c}_{L/2}).
\eea
We evolve the system unitarily with $\hat{H} = \hat{H}_0 + \hat{H}_C$ for a long time. Since $\hat{H}$ does not commute with $\hat{N}_r$, the particle number $N_r$ can change over time. To study \emph{how the left part influences the originally conserved particle number in the right part},
we calculate the charge variance of the right part,
\bea
    \delta N_r^2(t) = \bra{\Psi(t)} (\hat{N}_r-\bra{\Psi(t)} \hat{N}_r \ket{\Psi(t)})^2 \ket{\Psi(t)}.
\eea
The results for different system sizes $L$ are presented in Fig.~\ref{sfig3}. These results can be understood as follows: for eigenstates of $\hat{H}_0$ with the same total particle number but different particle numbers in the right part, $N_r$ and $N_r+1$, the energy difference is
\bea
    &-(2t_l \cos k_l-\mu_l)+(2t_r \cos k_r-\mu_r) \\
    =& (-2t_l \cos k_l+2t_r \cos k_r)+(\mu_l-\mu_r)\\
    \in & \big[-2(|t_l|+|t_r|)+(\mu_l-\mu_r),2(|t_l|+|t_r|)+(\mu_l-\mu_r)\big].
\eea
When $\mu_r<-2(|t_l|+|t_r|)+\mu_l$ or $\mu_r>2(|t_l|+|t_r|)+\mu_l$, the energy gap for a single pump exists, and particle transport from the left part cannot influence the particle number of the right part in the thermodynamic limit.

\section{Change of energy under periodic boundary kick}\label{app::r_2}
We evolve a state periodically as
\bea
    \ket{\Psi(t)} = U_F^{tf/2}\ket{\Psi(0)},
\eea
where the Floquet operator $U_F$ is
\bea\label{eq::U_F}
    U_F \equiv e^{-i\hat{H}_B/f}e^{-i\hat{H}_0/f}.
\eea
We study the energy difference
\bea
    \Delta E(t) = \bra{\Psi(t)}\hat{H} \ket{\Psi(t)} - \bra{\Psi(0)} \hat{H} \ket{\Psi(0)},
\eea
where $\hat{H} = \hat{H}_0 + \hat{H}_B$. In this work we restrict to a one-dimensional chain: $\hat H_0=\sum_j h_j$ is short-range with $\|h_j\|=O(1)$ and finite range $r_0$, and $\hat{H}_B$ has finite boundary support with $\|\hat H_B\|=O(1)$. All constants introduced below---except for the finite-size constant $\Lambda_L$---depend only on local energy scales and $r_0$, and are independent of the system size $L$.
Here, the operator norm $\|\cdot\|$ is the maximal absolute eigenvalue of the operator.

\subsection{Error Hamiltonian}

Define the error Hamiltonian $\hat{H}_{\mathrm{err}}$ via
\bea
    e^{-i\hat{H}_B/f}e^{-i\hat{H}_0/f} = e^{-i(\hat{H}_0+\hat{H}_B+\hat{H}_{\mathrm{err}})/f}.
\eea
Equivalently, with $A=-\frac{i}{f}\hat H_B$ and $B=-\frac{i}{f}\hat H_0$,
\bea
    \hat{H}_{\mathrm{err}} = if\Big[\log(e^A e^B)-(A+B)\Big].
\eea
By the Baker--Campbell--Hausdorff (BCH) expansion,
\bea
   \log(e^A e^B) \;=\; A+B+\tfrac12[A,B]
   +\tfrac{1}{12}\big([A,[A,B]]+[B,[B,A]]\big)+\cdots,
\eea
so $\hat H_{\mathrm{err}}$ is precisely the collection of higher-order commutator terms beyond $A+B$.

\subsection{Local growth of nested commutators}

For an operator $O$ acting nontrivially on only finitely many sites (bounded support), write ${\rm ad}_{\hat H_0}(O):=[\hat H_0,O]$ and ${\rm ad}_{\hat H_0}^m(O)$ for the $m$-fold commutator.

\paragraph{Key locality fact.}
Each commutator $[\hat H_0,\cdot]$ only involves those $h_j$ overlapping the current support, so the support can expand by at most $O(r_0)$ per step. Hence, after $m$ steps, the number of contributing local terms in $\hat H_0$ is bounded by $c_0+c_1 m$ with $c_0 = O(1)$ fixed by the initial support size of $O$ and $c_1=O(1)$ set by $r_0$.

\begin{lemma}[Generic 1D short-range]\label{lem:generic}
There exist $C,\Lambda=O(1)$ (set by local energy scales and $r_0$, independent of $L$) such that
\begin{equation}\label{eq:generic-bound}
  \big\|{\rm ad}_{\hat H_0}^{\,m}(O)\big\|
  \;\le\; C\,\Lambda^{m}\, m!\,\|O\|,
  \qquad m\ge 0.
\end{equation}
\end{lemma}
\begin{proof}
At step $k$ at most $c_0+c_1 k$ local terms touch the support; iterating $\|[\hat H_0,{\rm ad}_{\hat H_0}^{\,m-1}(O)]\| \le (c_0+c_1(m-1)) \|[h_j,{\rm ad}_{\hat H_0}^{\,m-1}(O)]\|\le 2 (c_0+c_1(m-1))\|h_j\|\|{\rm ad}_{\hat H_0}^{\,m-1}(O)\|$ produces a product $\prod_{k\le m}(c_0+c_1 k)\sim m!$.
\end{proof}

\begin{lemma}[Finite-size refinement in 1D]\label{lem:generic-finite}
On a finite chain of length $L$, there exist constants $C=O(1)$ and $\Lambda_L=O(L)$ (depending on local energy scales and $L$, but independent of $m$) such that for any local operator $O$ with bounded support
\begin{equation}\label{eq:generic-finite-bound}
  \big\|{\rm ad}_{\hat H_0}^{\,m}(O)\big\|
  \;\le\; C\,\Lambda_L^{m}\,\|O\|,
  \qquad m\ge 0.
\end{equation}
\end{lemma}
\begin{proof}
As in the proof of Lemma~\ref{lem:generic}, each commutator $[\hat H_0,\cdot]$ only involves those local terms $h_j$ that overlap the current support. On a finite chain of length $L$, the support of ${\rm ad}_{\hat H_0}^{\,m}(O)$ can never exceed the whole chain, so the number of contributing $h_j$'s is always bounded by $L$. Therefore, on a finite chain, the factorial $m!$ in \eqref{eq:generic-bound} is bounded by $(c_2 L)^m$ with $c_2 = O(1)$, and we have
\[
\big\|{\rm ad}_{\hat H_0}^{\,m}(O)\big\|
\;\le\; C(c_2\Lambda L)^m\,\|O\|.
\]
Thus \eqref{eq:generic-finite-bound} holds with $C=O(1)$ and $\Lambda_L= c_2 \Lambda L = O(L)$.
\end{proof}

\begin{lemma}[Free (quadratic) $H_0$]\label{lem:free}
If $\hat H_0$ is quadratic (free), then there exist $C,\lambda=O(1)$ such that
\begin{equation}\label{eq:free-bound}
  \big\|{\rm ad}_{\hat H_0}^{\,m}(O)\big\|
  \;\le\; C\,\lambda^{m}\,\|O\|,
  \qquad m\ge 0.
\end{equation}
\end{lemma}
\begin{proof}
Work in basis $\{M_\alpha\}$ of Majorana monomials,
expand $O=\sum_{\alpha\in S_\alpha} c_\alpha M_\alpha$. Here, $M_\alpha$ still act on only finitely many sites as $O$. Hence, $|S_\alpha|\le N_\alpha=O(1)$ and $\|c_\alpha M_\alpha\| \le C_*\|O\| (C_* =O(1))$.

For quadratic $H_0$, the commutator preserves operator size (number of Majoranas in a Majorana monomial). Moreover, locality of $\hat H_0$ and finiteness of local energy scales imply
\[
[\hat H_0,M_\alpha]=\sum_{\beta\in S_\beta(\alpha)} K_{\alpha\beta}\,M_\beta,
\quad |S_\beta(\alpha)|\le N_\beta = O(1),\quad |K_{\alpha\beta}|\le K_0 = O(1).
\]
Here, because the operator size of $M_\beta$ is the same as $M_\alpha$, $\|c_\alpha M_\beta\| \le C_*\|O\| (C_* =O(1))$. We get
\[
[\hat H_0,O] = \sum_{\alpha\in S_\alpha,\beta\in S_\beta(\alpha)}c_\alpha\, K_{\alpha\beta}\,M_\beta,
\]
and
\[
\|[\hat H_0,O]\| \le N_\alpha N_\beta K_0 C_* \,\|O\|
\]
Iterating this bound yields
\[
\big\|{\rm ad}_{\hat H_0}^{\,m}(O)\big\|
\;\le\; C\,\lambda^{m} \,\|O\|,
\qquad \lambda, C=O(1),
\]
which is a uniform geometric bound independent of $L$.
\end{proof}

\paragraph{Application to $A$ and $B$.}
Set $A:=-\tfrac{i}{f}\hat H_B$ and $B:=-\tfrac{i}{f}\hat H_0$. Then
\[
  {\rm ad}_{B}^{\,m}(A)=\Big(-\frac{i}{f}\Big)^{m+1}{\rm ad}_{\hat H_0}^{\,m}(\hat H_B).
\]
Using Lemma~\ref{lem:generic} (generic case), Lemma~\ref{lem:generic-finite} (finite-size refinement), and Lemma~\ref{lem:free} (free case), we obtain the following bounds:
\begin{align}
  \big\|{\rm ad}_{B}^{\,m}(A)\big\|
  &\;\le\; \frac{\|\hat H_B\|}{f^{m+1}}\,
           C\,\Lambda^{m}\, m!,
           &&\text{(generic 1D, thermodynamic limit)}\label{eq:ABB-generic}\\[4pt]
  \big\|{\rm ad}_{B}^{\,m}(A)\big\|
  &\;\le\; \frac{\|\hat H_B\|}{f^{m+1}}\,
           C\,\Lambda_L^{m},
           &&\text{(generic 1D, finite chain of length $L$)}\label{eq:ABB-generic-finite}\\[4pt]
  \big\|{\rm ad}_{B}^{\,m}(A)\big\|
  &\;\le\; \frac{\|\hat H_B\|}{f^{m+1}}\,
           C\,\lambda^{m},
           &&\text{(free $H_0$)}\label{eq:ABB-free}
\end{align}
where all constants are $O(1)$ in the thermodynamic limit, and $\Lambda_L=O(L)$ for finite $L$.

\medskip
\noindent\emph{Remarks.}
(i) The generic bound \eqref{eq:ABB-generic} is “exponential $\times$ factorial” and captures the linear growth of the light-cone; the factorial comes from multiplying $O(c_0+c_1 m)$ one-step factors.  
(ii) In the free case \eqref{eq:ABB-free}, the factorial disappears and one has a pure geometric bound.
(iii) On a finite chain \eqref{eq:ABB-generic-finite}, the factorial growth is cut off by the finite system size $L$ and can be replaced by a purely geometric bound $\propto L^{m}$.

\subsection{Bounding the BCH expansion}
Let $\mathrm{BCH}_n(A,B)$ be the homogeneous degree-$n$ part of
$\log(e^A e^B)$ ($n\ge2$). At each degree we can write
\[
  \mathrm{BCH}_n(A,B)=\sum_{i}c_i\,W_i^{(n)}(A,B),
\]
where $W_i^{(n)}(A,B)$ are independent nested commutators of total degree $n$ in $\{A,B\}$.

At degree $n$, Dynkin’s formula organizes terms into $k\ge1$ blocks of nested commutators~\cite{muger2019notes}.
A generic term carries a weight bounded by
\[
\frac{1}{m_1!\,n_1!\,\cdots\,m_k!\,n_k!},
\qquad m_1+n_1+\cdots+m_k+n_k=n,
\]
and we have at most $2^n$ degree-$n$ nested commutators in $\{A,B\}$. Thus
\bea\label{eq::bound_sum_num}
    \sum_{i} |c_i|
    \;\le\; \kappa^{\,n}\quad\text{for some }\kappa < 2 =O(1).
\eea

Every degree-$n$ nested commutator contains at least one $A$, and the worst-case norm is bounded by the pattern with exactly one $A$ and $(n-1)$ copies of $B$
\bea\label{eq::bound_each_term}
    \max_{i}\|W_i^{(n)}(A,B)\|
    \;\le\; \|\mathrm{ad}_B^{\,n-1}(A)\|
\eea

\paragraph{Generic 1D (no finite $f_c$ in the thermodynamic limit).}
Using \eqref{eq:ABB-generic}, \eqref{eq::bound_sum_num}, and \eqref{eq::bound_each_term}
\begin{equation}\label{eq:BCHn-generic}
  \|\mathrm{BCH}_n(A,B)\|
  \;\le\; \kappa^n \cdot \frac{\|\hat H_B\|}{f^n}\, C\,\Lambda^{\,n-1}\,(n-1)!
  \;=\; \frac{C\kappa\|\hat H_B\|}{f}\,\Big(\frac{\kappa\Lambda}{f}\Big)^{n-1}\,(n-1)!.
\end{equation}
Because of the $(n\!-\!1)!$ factor, the series $\sum_{n\ge2}\mathrm{BCH}_n(A,B)$ cannot converge for any fixed $f>0$ in the thermodynamic limit. Thus, for a generic 1D interacting system there is \emph{no} finite $f_c$ for norm convergence. Note that our argument is closely related to the proof of Lemma~1 in Ref.~\cite{kuwahara2016floquet}, which establishes an analogous bound for the Magnus expansion. The Magnus expansion is often referred to as the continuous analogue of the BCH formula~\cite{blanes2009magnus}.

\paragraph{Finite chains (finite $f_c(L)$ exists, dependent of $L$).}
On a finite chain of length $L$ we can instead use the finite-size bound \eqref{eq:ABB-generic-finite}, together with \eqref{eq::bound_sum_num} and \eqref{eq::bound_each_term}, to obtain
\begin{equation}\label{eq:BCHn-generic-finite}
  \|\mathrm{BCH}_n(A,B)\|
  \;\le\; \kappa^n \cdot \frac{\|\hat H_B\|}{f^n}\, C\,\Lambda_L^{\,n-1}
  \;=\; \frac{C\kappa\|\hat H_B\|}{f}\,\Big(\frac{\kappa\Lambda_L}{f}\Big)^{n-1}.
\end{equation}
This is a purely geometric bound in $n$ for any fixed $L$, so
$\sum_{n\ge2}\|\mathrm{BCH}_n(A,B)\|\ \text{converges if}\ f>f_c(L):=\kappa\Lambda_L$, where $f_c(L)$ depends on $L$.

\paragraph{Free (quadratic) $H_0$ (finite $f_c$ exists, independent of $L$).}
If $\hat H_0$ is quadratic, with \eqref{eq:ABB-free}, \eqref{eq::bound_sum_num}, and \eqref{eq::bound_each_term}
\begin{equation}\label{eq:BCHn-free}
  \|\mathrm{BCH}_n(A,B)\|
  \;\le\; \kappa^n \cdot \frac{\|\hat H_B\|}{f^n}\, C\,\lambda^{\,n-1}
  \;=\; \frac{C\kappa\|\hat H_B\|}{f}\,\Big(\frac{\kappa\lambda}{f}\Big)^{n-1}.
\end{equation}
This is a geometric bound, so $\sum_{n\ge2}\|\mathrm{BCH}_n(A,B)\|\ \text{converges if}\ f>f_c:=\kappa\lambda$.
Hence a finite convergence threshold exists in the free case, and $f_c$ does not depend on $L$.

\subsection{Bounding $\Delta E(t)$}
A single Floquet step is
\bea 
    U_F = e^{-i\hat H_B/f}e^{-i\hat H_0/f} = e^{-i(\hat H_0+\hat H_B + \hat H_{\mathrm{err}})/f} = e^{-i\hat H'/f},
\eea
where $\hat{H}' = \hat H_0+\hat H_B+\hat{H}_{\mathrm{err}}$.
The state at time $t$ is
\bea
    \ket{\Psi(t)} =& U_F^{tf/2}\ket{\Psi(0)}\\
    =& (e^{-i\hat{H}'/f})^{tf/2}\ket{\Psi(0)}\\
    =& e^{-i\hat{H}'t/2}\ket{\Psi(0)}.
\eea
Hence $\bra{\Psi(t)}\hat H'\ket{\Psi(t)}=\bra{\Psi(0)}\hat H'\ket{\Psi(0)}$.

Then
\bea\label{eq::bound_dE}
    |\Delta E(t)| =& |\bra{\Psi(t)}\hat{H} \ket{\Psi(t)} - \bra{\Psi(0)} \hat{H} \ket{\Psi(0)}|\\
    =& |(\bra{\Psi(t)}\hat{H} \ket{\Psi(t)} - \bra{\Psi(t)}\hat H'\ket{\Psi(t)}) - (\bra{\Psi(0)} \hat{H} \ket{\Psi(0)} - \bra{\Psi(0)}\hat H'\ket{\Psi(0)})|\\
    \le& |\bra{\Psi(t)}\hat{H} \ket{\Psi(t)} - \bra{\Psi(t)}\hat H'\ket{\Psi(t)}| + |\bra{\Psi(0)} \hat{H} \ket{\Psi(0)} - \bra{\Psi(0)}\hat H'\ket{\Psi(0)}|\\
    =& |\bra{\Psi(t)}\hat{H}_{\mathrm{err}}\ket{\Psi(t)}| + |\bra{\Psi(0)}\hat{H}_{\mathrm{err}}\ket{\Psi(0)}|\\
    \le& 2\|\hat{H}_{\mathrm{err}}\|.
\eea

Combining \eqref{eq:BCHn-generic}, \eqref{eq:BCHn-generic-finite}, and \eqref{eq:BCHn-free} with \eqref{eq::bound_dE}, we conclude:

\noindent\textbf{Generic 1D short-range.}
In the thermodynamic limit the norm $\|\hat{H}_{\mathrm{err}}\|$ obtained from the BCH series does not converge to a finite constant for any fixed $f>0$, and a generic interacting system can ev entually thermalize to infinite temperature. For any \emph{finite} system size $L$, however, the factorial growth $m!$ is cut off by $L$, see \eqref{eq:BCHn-generic-finite}. This yields an $L$-dependent threshold
\[
f_c(L)=\kappa\Lambda_L
\]
such that for $f>f_c(L)$ the BCH series converges, so that $|\Delta E(t)|\le 2\|\hat H_{\mathrm{err}}\|$ can remain small. In the numerical simulations presented in the main text we indeed observe that $|\Delta E|$ saturates to a small constant in the interacting system at high driving frequencies; this is precisely such a finite-size non-heating effect. As $L\to\infty$, $f_c(L)$ diverges and the non-heating window shrinks, consistent with eventual heating to infinite temperature in the thermodynamic limit~\cite{d2014long}.

\noindent\textbf{Free (quadratic) $\hat H_0$.}
There exists $f_c=O(1)$ such that for $f>f_c$ the BCH series converges
geometrically, hence
\[
\|\hat H_{\mathrm{err}}\|=O(1),\qquad
|\Delta E(t)|\le 2\|\hat H_{\mathrm{err}}\|=O(1),
\]
uniformly in system size.

\end{document}